\documentclass[11pt]{article}
\pdfoutput=1

\usepackage{bm}
\usepackage{epsfig}
\usepackage{amsmath,amsthm,amssymb,mathtools}
\usepackage{rotating}
\usepackage{float}
\usepackage{multirow}
\usepackage{array}
\usepackage{graphicx}
\usepackage{tikz}
\usetikzlibrary{patterns}
\tikzset{
  every picture/.style={line width=0.75pt},
  node/.style={rectangle, draw},
  ell/.style={ellipse, draw}
}

\usepackage{ulem}
\usepackage{natbib}         
\usepackage{color}
\usepackage{hyperref}      

\usepackage{epstopdf}
\makeatletter      
\newcommand{\rmnum}[1]{\romannumeral #1}
\newcommand{\Rmnum}[1]{\expandafter\@slowromancap\romannumeral #1@}
\makeatother

\usepackage[ruled]{algorithm2e}                 
\DeclareGraphicsExtensions{.eps}

\usepackage{tcolorbox}
\usepackage{multirow}
\usepackage{booktabs}
\usepackage{geometry}

\makeatletter
\def\singlespace{\def\baselinestretch{1}\@normalsize}

\@addtoreset{equation}{section}
\renewcommand{\baselinestretch}{1.412}

\newtheorem{lemma}{{\bf Lemma}}
\newtheorem{remark}{{\bf Remark}}
\newtheorem{theorem}{{\bf Theorem}}
\newtheorem{Condition}{{\bf Condition}}

\newcommand{\wh}{\widehat}
\newcommand{\wt}{\widetilde}

\newcommand{\diag}{\mbox{diag}}

\newcommand{\T}{\!\mbox{\scriptsize T}}
\newcommand{\var}{\mbox{var}}

\newcommand{\bbeta}{\mbox{\boldmath{$\beta$}}}

\newcommand{\btheta}{\mbox{\boldmath{$\theta$}}}

\newcommand{\bI}{\mbox{\bf I}}

\newcommand{\bX}{\mbox{\bf X}}

\newcommand{\bB}{\mbox{\bf B}}
\newcommand{\bC}{\mbox{\bf C}}
\newcommand{\bS}{\mbox{\bf S}}

\newcommand{\bW}{\mbox{\bf W}}

\newcommand{\bP}{\mbox{\bf P}}

\newcommand{\bU}{\mbox{\bf U}}

\newcommand{\bZ}{\mbox{\bf Z}}

\newcommand{\bSigma}{\boldsymbol{\Sigma}}

\definecolor{darkblue}{RGB}{0,0,169}
\definecolor{skyblue}{RGB}{135,206,255}
\definecolor{orange}{RGB}{255,165,0}

\LetLtxMacro{\@stdmakecaption}{\@makecaption}
\renewcommand{\@makecaption}[2]{\@stdmakecaption{#1}{#2}}   

\begin{document}
\title{\bf
Stab-GKnock: Controlled variable selection for partially linear models using generalized knockoffs
      }

\author{
Han Su, Panxu Yuan, Qingyang Sun, Mengxi Yi\thanks{Corresponding author, Email: \texttt{mxyi@bnu.edu.cn}.}, Gaorong Li
\\
School of Statistics, Beijing Normal University, \\
Beijing 100875, P. R. China
}

\maketitle

\begin{abstract}

The recently proposed fixed-X knockoff is a powerful variable selection procedure that controls the false discovery rate (FDR) in any finite-sample setting, yet its theoretical insights are difficult {to show} beyond Gaussian linear models.
In this paper, we make the first attempt to extend the fixed-X knockoff to partially linear models by using generalized knockoff features, and propose a new stability generalized knockoff (Stab-GKnock) procedure by incorporating selection probability as feature importance score.
We provide FDR control and power guarantee under some regularity conditions.
In addition, we propose a two-stage method under high dimensionality by introducing a new joint feature screening procedure, {with guaranteed sure screening property}.
Extensive simulation studies are conducted to evaluate the finite-sample performance of the proposed method. 
A real data example is also provided for illustration.

\vspace{0.2in}

\end{abstract}

\section{Introduction}
\label{intro}
Semiparametric regression models have been widely used to balance between modeling bias and ``curse of dimensionality'' for modeling complex data in many scientific fields, including information sciences,
econometrics,
biomedicine, social sciences, and so on.
See the monographs \citep{RWC2003,HMSW2004,Xue2012,LZF2016} 
for more details.
As the leading example of semiparametric models, partially linear models (PLM) \citep{EGRW1986,HLG2000} 
hold both the flexibility of nonparametric models and model interpretation of linear models. 
Specifically, 
PLM takes the form
\begin{equation}
\label{model1}
	Y={\bm X}^{\rm T}{\bm \beta}+g(U)+\varepsilon,
\end{equation}
where $Y\in \mathbb{R}$ is a response variable, ${\bm{X}}=(X_{1},\dots,X_{p})^{\T} \in \mathbb{R}^p$ is an explanatory covariate vector, $\bm{\beta}=({\beta}_1,\dots,{\beta}_p)^{\T}$ is a \textit{p}-dimensional vector of unknown regression coefficients, $U$ is an observed univariate variable, $g(\cdot)$ is an unknown smooth function, 
$\varepsilon \sim N(0,\sigma^2)$ with $0<{\sigma}^{2}<\infty$, and independent of the associated covariates $({\bm{X}}^{\T},U)$.

Variable selection for high-dimensional PLM has attracted extensive attention over the past two decades. 
When the dimension $p$ of the linear part diverges slowly with the sample size $n$,
\cite{XH2009} 
proposed SCAD-penalized estimators of the linear coefficients and established the consistency results. 
\cite{WXQL2014}
proposed a doubly penalized procedure to identify significant linear and nonparametric additive components.
Allowing $p>n$ or even to grow exponentially with $n$,
\cite{LWT2012} 
proposed the profile forward regression (PFR) algorithm to perform feature screening for ultra-high-dimensional PLM. 
\cite{Zhu2017} 
proposed a new two-step procedure for estimation and variable selection.  
\cite{LZL2019} 
proposed the projected estimation for massive data and established consistency results for the linear and nonparametric components.
For more variable selection methods in PLM, please refer to  \cite{Bunea2004}, \cite{LL2009}, \cite{LZXF2010}, \cite{WXQL2014}, \cite{LLLT2017}, \cite{LLT2017}, \cite{LL2022}.
Yet, most existing methods in the literature mainly focus on how to select all significant variables, lacking adequate attention to the control of selection error rates such as false discovery rate (FDR).

Loosely speaking, FDR is defined as the expectation of the proportion of false discoveries among all discoveries, which was first introduced in \cite{BH1995} and since then, has been a gold criterion in large-scale multiple testing.
However, traditional FDR control methods, 
such as \cite{BY2001}, \cite{Storey2002}, \cite{Efron2007}, \cite{FF2011}, \cite{SC2016}, \cite{JJ2019}, \cite{MCL2021}, \cite{FL2022} and among others, 
rely heavily on p-values as feature important measures.
This limits the application to high-dimensional PLM analysis, partially due to non-negligible estimate bias introduced by the nonparametric component $g(U)$ and the regularized terms, which makes p-values difficult to obtain \citep{ZYC2019}.

More recently, \cite{BC2015} 
proposed an elegant fixed-X knockoff procedure under low-dimensional Gaussian linear models to achieve FDR control without resorting to p-values.
The main point is to generate ``fake'' knockoff features that mimic the dependency structure of the original variables.
The application of fixed-X knockoff has been investigated in many aspects.
\cite{BC2019} examined the performance of fixed-X knockoff when $p>n$, and proposed a ``screening+knockoff'' two-stage procedure for high dimensional setting based on data splitting technique.
\cite{LSK2022} introduced generalized knockoff features for the structural change detection, and achieved FDR control under the dependent structure.
See more details in 
\cite{DB2016}, \cite{SXZ2020}, \cite{LM2021}, 
\cite{YFL2022}, \cite{CSY2023}
and references therein. 
However, one common feature of existing works based on fixed-X knockoff is that they actually focus on the linear regression setting.
When extending beyond Gaussian linear models, the failure of the sign-flip property for knockoff statistics \citep{BC2015,CFJL2018} renders the FDR control infeasible. Despite the fact that \cite{LSK2022} and \cite{CSY2023} 
have made some relevant pioneering discussions, they only focused on the structural change detection problem in sparse linear regressions. 
In addition, there is generally a lack of theoretical power analysis for knockoff-based selection in semiparametric models, except for \cite{FDLL2020} and \cite{DLL2022}, who have addressed this issue from a simulation perspective or in nonparametric additive models.
Nevertheless, both of them are introduced based on the model-X knockoff framework in \cite{CFJL2018}, which is a randomized procedure and counts heavily on the knockoff features generation mechanism.
Whereas, it remains open for the fixed-X knockoff based selection for semiparametric models.

In this paper, we re-study the selection probability statistics \citep{YFL2022,YKL2023} and propose a novel stability generalized knockoffs (Stab-GKnock) procedure to study FDR control and power analysis for PLM (\ref{model1}).
To the best of our knowledge, this is the first attempt to extend the fixed-X knockoff beyond Gaussian linear models and study the problem of controlled variable selection for semiparametric partially linear models. 
We emphasize that this extension is not trivial since the presence of the nonparametric part $g(U)$ makes the sign-flip property of knockoff statistics difficult to verify, which will be further discussed in Sections \ref{pre} and \ref{metho}. 
Three key components are summarized for our implementation: 
the construction of generalized knockoff features, 
the intersection subsampling-selection strategy,
and the two-stage extension with a new joint feature screening method.

The workflow and algorithms are presented in Figure \ref{workflow}, Algorithms \ref{algo1} and  \ref{algo2}, respectively. 
Specifically, we first apply the projection technique to recover the active set and transform the original data. 
We then construct the generalized knockoff features based on the projected design matrix and establish two pairwise exchangeability properties with the dependent projected data. 
Noting that the traditional Lasso signed max (LSM) statistic does not perform well facing high correlation structure \citep{DGSZ2023}, 
we innovate the idea of stability selection \citep{MB2010,SS2013} based on an intersection subsampling-selection strategy, and incorporate selection probability difference (SPD) as generalized knockoff statistics, see Figures \ref{scatter plot} and \ref{histogram} in Section \ref{metho}. 
We theoretically show the proposed Stab-GKnock procedure achieves the finite-sample FDR control and asymptotic power one. 
To extend the applications to high-dimensional PLM, we further propose a two-stage procedure using data splitting technique, which are commonly considered in knockoff-based literature, such as \cite{BC2019}, \cite{FDLL2020}, \cite{SXZ2020}, \cite{LKLL2022}
and among others. 
In the first step, we use the first part of data to reduce the dimension to a suitable order by introducing a new joint screening method called sparsity constrained projected least squares (Sparse-PLS, SPLS) method. 
In contrast with the traditional marginal effect screening, Sparse-PLS naturally accounts for the joint effects among features and performs better in applications.
In the second step, we apply Stab-GKnock to select variables on the screened variables set using the second part of data.
Theoretical analysis shows the Sparse-PLS screening method enjoys the sure screening property.
The theoretical guarantees in terms of both FDR and power are also established. 

The rest of this paper is organized as follows. 
We begin in Section \ref{pre} with a brief review of the fixed-X knockoff framework, as well as a detailed description of the model setting and the projection technique. 
We propose the methodology of the Stab-GKnock procedure by constructing the generalized knockoff features and the refined SPD statistics in Section \ref{metho}. The associated theoretical results in terms of both FDR and power are also established under some regularity conditions. 
Section \ref{HD-setting} proposes the Sparse-PLS screening method and shows its sure screening property under some mild conditions, and further studies a two-stage extension with FDR control for the high-dimensional setting. 
Section \ref{simu} assesses the finite sample performance of the Stab-GKnock, Sparse-PLS, and SPLS-Stab-GKnock procedure with several simulation studies. 
A real data application is also provided in Section \ref{real}. 
We briefly summarize this paper in Section \ref{Conclusion}. 
Technique proofs of theories and additional simulation studies are provided in Supplementary material.

\section{Preliminaries}
\label{pre}
To avoid confusion, we first specify some notations to facilitate the presentation. 
The boldface roman $\bf B$ represents a matrix, and the boldface italics $\bm B$ represents a vector. For a subset $\mathcal{A} \subset {\{1,\dots,p\}}$, we denote $|\mathcal{A}|$ and $\mathcal{A}^c$ as its cardinality and complement set, respectively. 
For a $n \times p$ matrix $\bf B$ and a generic set $\mathcal{A}$, we use 
${\bf B}_{\mathcal{A}}=\{b_{ij},i=1,\dots,n, j\in \mathcal{A}\} \in \mathbb{R}^{n\times|\mathcal{A}|}$ to represent the submatrix consisting of the column of $\bf B$ with indices in $\mathcal{A}$, and ${\bm B}_{\mathcal{A}}=\{b_{j},j \in \mathcal{A}\} \in \mathbb{R}^{|\mathcal{A}|}$ to represent the subvector of $\bm B$ corresponding to $\mathcal{A}$. 
Similarly, for a subset $I \in {\{1,\dots,n\}}$, we use 
${\bf B}(I)=\{b_{ij}, i \in I, j=1,\dots,p\} \in \mathbb{R}^{|I|\times p}$ to represent the submatrix consisting of the row of $\bf B$ with indices in $I$. 
Denote $\lambda_{\min}(\bf{B})$ and $\lambda_{\max}(\bf B)$ the smallest and largest eigenvalues of an arbitrary square matrix $\bf B$, respectively.
For two constants $a$ and $b$, let $a\vee b$ and $a\wedge b$ be the maximum and minimum between $a$ and $b$.
For a matrix $\bB=\{b_{ij}\}$, let 
$\|\bB\|_{1} = \max\limits_j(\sum\limits_i |b_{ij}|), 
\|\bB\|_{2} = \sqrt{\lambda_{\max}(\bf{B}^{\T}\bf{B})}, 
\|\bB\|_\infty = \max\limits_i(\sum\limits_j |b_{ij}|), 
\|\bB\|_\text{max} = \max\limits_{i,j}(|b_{ij}|).$
For a vector $\bm{B}=(b_1,\dots,b_p)^{\T}$, we denote
$\|\bm{B} \|_{\infty}=\max\limits_{1\leq i\leq p} |b_i|$,
and $\|\bm{B} \|_q=(\sum\limits_{i} |b_i|^q)^{1/q}$ the $L_q$-norm for $q\in (0,\infty)$.
Let $\lfloor \cdot\rfloor$ be the floor function.
Let $\bm{1} (\cdot)$ denote the indicator function,
$\bI_n$ the $n \times n$ identity matrix,
$\bm{e}_j=(0,\dots,0,1,0,\dots,0)^{\T}$ the vector with the \textit{j}-th component equals to 1 while the other components equal to 0.
{$\bB \succeq 0$ denotes $\bB$ a positive semidefinite matrix,
$a_n \asymp b_n$ denotes sequences $\{a_n\}$ and $\{b_n\}$ have the same order of magnitude}.
Throughout the paper, we use $c, C,\dots$ to denote constants that may vary from place to place.

\subsection{{Review of fixed-X knockoff}}
\label{section2.1}
Knockoffs methods are a flexible class of reproducible multiple testing procedures with FDR control.
The original fixed-X knockoff filter in \cite{BC2015} 
considers the classical Gaussian linear model
\begin{equation*}
	\bm{Y}={\bX}{\bm \beta}+\bm{\varepsilon},\qquad \bm{\varepsilon} \sim N(\bm{0},\sigma_{\varepsilon}^2 \bI_n),
\end{equation*}
where $\bm{Y}\in \mathbb{R}^{n}$ is the response, 
$\bX \in \mathbb{R}^{n\times p}$ is the fixed design matrix, $0<\sigma_{\varepsilon}^2<\infty$ and $\bbeta\in \mathbb{R}^{p}$ are unknown. 
The goal is to test the $p$ hypotheses $\mathrm{H}_{0 j}: \beta_{j} = 0$ against the two-sided alternative, for $j=1,\dots,p.$
The valid knockoffs $\wh{\bX}=(\wh{\bm X}_1,\dots,\wh{\bm X}_p)$ for ${\bX}$ can be constructed, obeying that
\begin{equation}
\label{classical knockoffs}
    {\wh\bX}^{\T}{\wh\bX}={\bX}^{\T}{\bX}, \qquad 
    {\wh\bX}^{\T}{\bX}={\bX}^{\T}{\bX}-\diag\{\bm{s}\}
\end{equation}
for some $\diag\{\bm{s}\}$ satisfying $2\bSigma-\diag\{\bm{s}\} \succeq 0$, where $\bm s=(s_1,\ldots,s_p)^{\T} \in \mathbb{R}^p_+.$ When $n \geq 2p$, an explicit representation can be computed by 
$\wh\bX = {\bX}\left[\bI_p-({\bX}^{\T}{\bX})^{-1} \diag\{\bm{s}\}\right]+\wh{\bU}\bC_0$, where $\wh{\bU} \in \mathbb{R}^{n\times p}$ is an orthonormal matrix orthogonal to the column space of $\bX$, $\bC_0$ is the Cholesky decomposition factor of the matrix $2\diag\{\bm{s}\}-\diag\{\bm{s}\} ({\bX}^{\T}{\bX})^{-1} \diag\{\bm{s}\}$,
see the details in \cite{BC2015}. 
Once $\wh\bX$ is constructed, knockoffs framework then calculates the knockoff statistics ${\bm W}=(W_1,\dots,W_p)^{\rm T}$ based on $([\bX,\wh{\bX}],\bm{Y})$ obeying the following two properties:
\begin{description}
	\item[(\rmnum{1}) (The sufficient property).] The knockoff statistics $\bm W$ only depend on the augment Gram matrix $[\bX,\wh{\bX}]^{\rm T} [\bX,\wh{\bX}]$ and the feature-response inner products $[\bX,\wh{\bX}]^{\rm T} \bm{Y}.$

	\item[(\rmnum{2}) (The antisymmetry property).] Swapping the $j$-th column of $\bX$ with the associated knockoff counterpart, it only changes the sign of the knockoff statistic $W_j$, i.e., for any $\mathcal{A} \subset \{1,\dots,p\}, j=1,\dots,p,$
    \begin{equation}
    \label{antisymmetry}
        W_{j}\left([\bX,\wh{\bX}]_{{\rm swap}(\mathcal{A})}, \bm{Y}\right)=
        W_{j}\left([\bX,\wh{\bX}], \bm{Y}\right) \cdot
        \begin{cases}
		&+1,\quad \text{if}\enspace j \notin \cal{A},\\
		&-1,\quad \text{if}\enspace j \in \cal{A},
	\end{cases}
    \end{equation}
    where ${\rm swap}(\mathcal{A})$ is an operator that swaps $\bX_{\mathcal{A}}$ and $\wh{\bX}_{\mathcal{A}}$.
\end{description}
The type of knockoff statistic is not unique. 
Knockoff demands statistics possessing the sign-flip property \citep{BC2015,CFJL2018} 
as follows
\begin{equation} 
\label{sign-flip}
(W_1,\dots,W_p) \overset{d}{=} (\epsilon_1 \cdot W_1,\dots, \epsilon_p \cdot W_p),
\end{equation}
where $\epsilon_j=1 ~\text{if}~ \beta_{j} = 0, ~\text{and}~ \epsilon_j = \pm 1$ with equal probability $1/2$ if $\beta_{j} \neq 0$. 
The sign-flip property is key to obtain valid error control in the knockoff framework, 
we will further explain the details in Section \ref{asy1}.
The sign-flip property is a consequence of following two exchangeability properties for $([\bX,\wh{\bX}],\bm{Y})$.
\begin{description}
	\item[(\rmnum{3}) (Pairwise exchangeability for the features).] For any subset $\mathcal{A} \subset \{1,\dots,p\}$, we have
        \begin{equation}
        \label{exchangeability1}
            [\bX,\wh{\bX}]^{\rm T}_{{\rm swap}(\mathcal{A})} [\bX,\wh{\bX}]_{{\rm swap}(\mathcal{A})}=
            [\bX,\wh{\bX}]^{\rm T} [\bX,\wh{\bX}].
        \end{equation}

	\item[(\rmnum{4}) (Pairwise exchangeability for the response).] For any subset $\mathcal{G} \subset \{j: \beta_j = 0\}$, we have
        \begin{equation}
        \label{exchangeability2}
            [\bX,\wh{\bX}]^{\rm T}_{{\rm swap}(\mathcal{G})} \bm{Y}
            \overset{d}{=}
            [\bX,\wh{\bX}]^{\rm T} \bm{Y},
        \end{equation}
\end{description}
where the property (\ref{exchangeability2}) demands the i.i.d. structure for the response $\bm{Y}$.
After calculating knockoff statistics, 
the fixed-X knockoff filter rejects $\mathrm{H}_{0 j}$ if $W_j \geq T$, where $T$ is a data-dependent threshold.
There are two ways to choose the threshold $T$, one is defined as
\begin{equation}
\label{threshold1}
    T=\min\left\{t \in \mathcal{W}
    :\frac{\left|\{j:W_j<-t\}\right|} {\left|\{j:W_j>t\}\right|\vee 1}\le q\right\},\quad(\text{Knockoff})
\end{equation}
another is defined as
\begin{equation}
\label{threshold2}
    T=\min\left\{t \in \mathcal{W}
    :\frac{1+\left|\{j:W_j<-t\}\right|} {\left|\{j:W_j>t\}\right|\vee 1} \le q\right\}, \quad(\text{Knockoff+})
\end{equation}
where $\mathcal{W} \coloneqq \{|W_j|:|W_j|>0\}$. The threshold $T$ chosen by Knockoff+ (\ref{threshold2}) is slightly more conservative than the threshold chosen by Knockoff (\ref{threshold1}), and satisfies FDR control in finite-sample setting. Intuitive extensions to $p<n<2p$ have also been given in \cite{BC2015}.

\subsection{{Recap: Projected spline estimation in PLM}}
\label{section2.2}
{Suppose $\{(\bm{x}_{i}^{\T},U_{i},Y_{i}),i=1,\dots,n\}$ are observed samples of $(\bm{X}^{\T},U,Y)$ from model (\ref{model1}), 
then model (\ref{model1}) can be re-expressed with matrix form}
\begin{equation}
\label{model2}
	\bm{Y}={\bX}{\bm \beta}+g(\bm{U})+\bm{\varepsilon},
\end{equation}
where  $\bm{Y}=(Y_1,\dots,Y_n)^{\T} \in \mathbb{R}^{n}$ is the response vector, $\bX=(\bm{x}_1,\dots,\bm{x}_n)^{\T}=(\bm{X}_1,\dots,\bm{X}_p) \in \mathbb{R}^{n\times p}$ is the fixed design matrix with $\bm{x}_i=(X_{i1},\dots,X_{ip})^{\T}$ and $\bm{X}_j=(X_{1j},\dots,X_{nj})^{\T}$, $\bm{U}=(U_1,\dots,U_n)^{\T} \in \mathbb{R}^{n}, g(\bm{U})=(g(U_1),\dots,g(U_n))^{\T}$, and $\bm{\varepsilon}=(\varepsilon_1,\dots,\varepsilon_n)^{\T}$ is the vector of model error. 

In this subsection, we will introduce the projected spline estimators of $\bbeta$ and $g(\cdot)$ in the partially linear model (\ref{model2}). Specifically, we use the polynomial splines to approximate the nonparametric part $g(\cdot)$ which satisfies certain smoothness.
According to the splines’ approximation property \citep{de2001,Huang2003,Schumaker2007}, 
the nonparametric function in model (\ref{model2}) can be well approximated and parameterized as $g(\bm{U}) = (g(U_1),\dots,g(U_n))^{\T} \approx \bZ\btheta_{0}$, where $\btheta_0 \in \mathbb R^K$ is an unknown parametric vector, $\bZ=(\bm{B}(U_{1}),\dots,\bm{B}(U_{n}))^{\T} \in \mathbb R^{n \times K}$ is the known basis matrix. $\bm{B}(u)=(B_{1}(u), \dots ,B_{K}(u))^{\T}$ is the B-spline basis {with $K=K^*+m$, where $K^*$ is the number of internal knots, and $m$ is the order of polynomial splines.} 
Thus, the problem of estimating $g(\cdot)$ becomes that of estimating $\btheta_0$.
We consider the following penalized least squares objective function
\begin{equation}
\label{2.3}
\mathcal{Q}(\bbeta,\btheta_0)=
\frac{1}{2} \|\bm{Y}-\bX \bbeta-\bZ \btheta_0\|^{2}+ \lambda \|\bbeta \|_{1},
\end{equation}
where $\lambda \geq 0$ is a tuning parameter. We then adopt the projection technique to transfer (\ref{2.3}) to a Lasso-type problem \citep{Tibshirani1996}. 
For any given $\bbeta$, a minimizer $\wt\btheta$ of $\mathcal{Q}(\bbeta,\btheta_0)$ is 
defined as
\begin{equation}\label{2.4}
    \wt\btheta=(\bZ^{\T}\bZ)^{-1}\bZ^{\T}(\bm{Y}-\bX\bbeta).
\end{equation}

Let $\bP_{\bf Z}=\bZ(\bZ^{\T}\bZ)^{-1}\bZ^{\T}$ be the projection matrix of the column space of the basis matrix $\bZ$, $\bW \coloneqq \bI_n-\bP_{\bf Z}$ is also a symmetric idempotent matrix. For simplicity, let $\bm{Y}^*=\bW\bm{Y}, \bX^*=\bW\bX$. 
By (\ref{2.3}), (\ref{2.4}) and some simple calculations, we can obtain the projected spline estimator of $\bbeta$ as
\begin{equation}\label{2.5}
    \wh{\bbeta}=
    \arg \min_{\bm{\beta}}\left\{ \frac{1}{2n} \|\bW(\bm{Y}-\bX \bbeta)\|^{2}+ \lambda \|\bbeta \|_{1} \right\} = 
    \arg \min_{\bm{\beta}} \left\{ \frac{1}{2n} \|\bm{Y}^*-\bX^*\bbeta\|^{2}+ \lambda \|\bbeta \|_{1}\right\}.
\end{equation}
After obtaining $\wh\bbeta$, we can plug it back into (\ref{2.4}) to obtain
\begin{align}
	\label{2.6}\notag
	\wh{\btheta}&=(\bZ^{\T}\bZ)^{-1}\bZ^{\T}(\bm{Y}-\bX\wh{\bbeta}),\\
	\wh{g}&=\bZ(\bZ^{\T}\bZ)^{-1}\bZ^{\T}(\bm{Y}-\bX\wh{\bbeta})=\bP_{\bf Z}(\bm{Y}-\bX\wh{\bbeta}).
\end{align}

The oracle inequalities and sign consistency of the projected spline estimator $\wh{\bbeta}$ have been established in the literature \citep{MH2015,LZL2019}.
Nonetheless, the finite-sample FDR control is of more interest for researchers, yet faces severe challenges.

\subsection{Problem setup}
\label{section2.3}

{In this paper}, we extend the knockoffs framework to the partially linear model (\ref{model1}), aiming to select as many truly associated variables as possible while keeping FDR at a predetermined level. 
Denote $[p]=\{1,\dots,p\}$ the index set of the full model, and $\mathcal{S}=\{j:\beta_j\neq0 \}$ the active set, i.e., the index set of non-null features. 
$\mathcal{S}^c=[p] \setminus \mathcal{S}$ is the unactive set. $p_1 \coloneqq |\mathcal{S}|$ and $p_0 \coloneqq |\mathcal{S}^c|= p-p_1$ are the numbers of the relevant and null features.  
Let $\widehat{\mathcal{S}}$ denote the discovered variable set by some variable selection procedures, FDR and power of a variable selection procedure are defined, respectively, as

\begin{equation*}
{\rm FDR}(\widehat{\mathcal{S}}) = \mathbb{E}\left[\frac{|\widehat{\mathcal{S}} \cap \mathcal{S}^c|}{|\widehat{\mathcal{S}}| \vee 1}\right],\qquad
{\rm Power}(\widehat{\mathcal{S}}) = \mathbb{E}\left[\frac{|\widehat{\mathcal{S}} \cap \mathcal{S}|}{|\mathcal{S}|} \right].
\end{equation*}

The idea of extending knockoff framework to PLM is intuitive, but not trivial when constructing the knockoff features. On account of the randomness of $\bm{U}$, classical knockoff features (\ref{classical knockoffs}) based on $\bX$ will lead the property (\ref{exchangeability2}) fail. 
Therefore, we consider to construct knockoff features based on the transformed design matrix $\bX^*$. Note that $\bX^*=\bW\bX$ is associated with the projection matrix $\bW$, the elements of the transformed response $\bm{Y}^*=\bW\bm{Y}$ are no longer i.i.d. since
\begin{equation}\label{transformed cdf}
    \bm{Y}^* \sim N(\bX^*\bbeta, \sigma^2\bW).
\end{equation}
The above dependence structure violates the assumption imposed for the fixed-X knockoff in \cite{BC2015}, 
further makes the sign-flip property (\ref{sign-flip}) difficult to verify, hence calls new methodological and theoretical investigations. 
In Section \ref{metho}, we will develop a novel Stab-GKnock procedure and establish the desired FDR and power guarantee for partially linear models.

\section{Methodology}
\label{metho}



\subsection{Extending the fixed-X knockoff to PLM: Stab-GKnock}
\label{sec Stab-GKnock}
In this subsection, we extend the fixed-X knockoff to partially linear models using generalized knockoff features in \cite{LSK2022}. 
To conduct FDR control and power analysis, 
we provide a nontrivial technical analysis to prove the pairwise exchangeability properties for the dependent transformed data. 
We also innovate the idea of stability selection in \cite{MB2010} 
based on an intersection subsampling-selection strategy, and incorporate the selection probability difference as generalized knockoff statistics. 
Hence, we call this procedure as Stab-GKnock, which can be concluded in following three steps.

\vskip4mm
\noindent 
{\bf Step 1: Construct generalized knockoff features.}
For simplicity, let $\bX^*=(\bm{x}^*_1,\dots,\bm{x}^*_n)^{\T}=(\bm{X}^*_1,\dots,\bm{X}^*_p) \in \mathbb{R}^{n \times p}, \bm{Y}^*=(Y^*_1,\dots,Y^*_n)^{\T} \in \mathbb{R}^{n}.$
Without loss of generality, we assume that the transformed design matrix $\bX^*$ is standardized such that $\|\bm{X}^*_j\|_2^2=1, j=1,\dots,n.$ Denote $\bSigma^*={\bX^*}^{\T}\bX^*$ the Gram matrix of $\bX^*$. 
Then we construct the generalized knockoff features $\wt\bX=(\wt{\bm{X}}_1,\dots,\wt{\bm{X}}_p) \in \mathbb{R}^{n \times p}$ based on $\bX^*$ instead of original design $\bX$, satisfing
\begin{equation}\label{GKnockoff1}
    {\wt\bX}^{\T}{\wt\bX}=\bSigma^*, \qquad
    {\wt\bX}^{\T}{\bX^*}=\bSigma^*-\diag\{\bm{s}\},
\end{equation}
where $\bm{s}=(s_1,\dots,s_p)^{\T} \in \mathbb{R}^p_+.$ The generalized knockoff matrix $\wt\bX$ mimics the dependency structure of the transformed design $\bX^*$, see Section \ref{section2.1}. When $n \geq 2p$, one can compute $\wt\bX$ by
\begin{equation}\label{GKnockoff2}
    \wt\bX = {\bX^*}\left[\bI_p-(\bSigma^*)^{-1} \diag\{\bm{s}\}\right]+\wt{\bU}\bC.
\end{equation}
Here, $\diag\{\bm{s}\}$ is a diagonal matrix obeying $2\bSigma^*-\diag\{\bm{s}\} \succeq 0$, 
$\wt{\bU} \in \mathbb{R}^{n\times p}$ is an orthonormal matrix that is orthogonal to the column space of $\bX^*$, i.e., ${\wt{\bU}^{\T}{\bX^*}}={\bf 0}$, 
and $\bC$ is the Cholesky decomposition factor of the matrix $2\diag\{\bm{s}\}-\diag\{\bm{s}\} (\bSigma^*)^{-1} \diag\{\bm{s}\}.$

\begin{remark}
    The projection transformation does not affect the existence of generalized knockoff features $\wt\bX$.
    According to (\ref{GKnockoff2}), $\wh\bX$ exists if and only if $\bSigma^*$ is reversible. 
    By the definition of the B-spline basis $\bm{B}(u)$ and the basis matrix $\bZ$, $\bZ$ is of full column rank $K$, 
    {where $K$ is the dimension of B-spline basis}.
    This implies that the projection matrix $\bW=\bI_n-\bP_{\bf Z}$ is of rank $n-K \geq p$ if $K\leq p$.
    Thus, $\bX^*=\bW\bX$ is of full column rank $p$ and $\bSigma^*$ is invertible. 
\end{remark}



In the proposed Stab-GKnock procedure, we make the first attempt to establish two pairwise exchangeability properties for PLM, shown in Theorems \ref{th1} and \ref{th2}. As we have emphasized in Section \ref{section2.1}, they are essential for the sign-flip property (\ref{sign-flip}) of statistics, yet not trivial since the elements of the transformed response $\bm{Y}^*=\bW\bm{Y}$ are no longer i.i.d.

\begin{theorem}[Pairwise exchangeability for the generalized knockoff features]
\label{th1}
For any subset $\mathcal{A} \subset \{1,\dots,p\}$, we have
\begin{equation*}
\label{exchangeability3}
    [\bX^*,\wt{\bX}]^{\rm T}_{{\rm swap}(\mathcal{A})} [\bX^*,\wt{\bX}]_{{\rm swap}(\mathcal{A})}=
    [\bX^*,\wt{\bX}]^{\rm T} [\bX^*,\wt{\bX}].
\end{equation*}
\end{theorem}

\begin{theorem}[Pairwise exchangeability for the transformed response]
\label{th2}
For any subset $\mathcal{G} \subset \mathcal{S}^c$, we have
\begin{equation*}
\label{exchangeability4}
    [\bX^*,\wt{\bX}]^{\rm T}_{{\rm swap}(\mathcal{G})} \bm{Y}^*
    \overset{d}{=}
    [\bX^*,\wt{\bX}]^{\rm T} \bm{Y}^*.
\end{equation*}
\end{theorem}

Theorem \ref{th1} shows that the Gram matrix of $[\bX^*,\wt{\bX}]$ is invariant when we swap the $j$-th column of $\bX^*$ and $\wt{\bX}$ for each $j \in \mathcal{A}.$ 
{The proof of Theorem \ref{th1} is referred to Supplementary material S.2.}
Theorem \ref{th2} shows that the distribution of the inner product $[\bX^*,\wt{\bX}]^{\rm T} \bm{Y}^*$ is invariant when we swap the $j$-th column of $\bX^*$ and $\wt{\bX}$ for each $j \in \mathcal{G}.$ 
According to (\ref{transformed cdf}), we can obtain the swapped distribution as follows
\begin{equation*}
\label{swapped cdf}
    [\bX^*,\wt{\bX}]^{\rm T}_{{\rm swap}(\mathcal{G})} \bm{Y}^*
    \sim N([\bX^*,\wt{\bX}]^{\rm T}_{{\rm swap}(\mathcal{G})} \bX^*\bbeta, 
    \sigma^2[\bX^*,\wt{\bX}]^{\rm T}_{{\rm swap}(\mathcal{G})} \bW [\bX^*,\wt{\bX}]_{{\rm swap}(\mathcal{G})}).
\end{equation*}
Under the Gaussian assumption, Theorem \ref{th2} holds if and only if the expectation and covariance matrix of the swapped distribution are invariant. The expectation is invariant as $\beta_j=0, j \in \mathcal{G}.$ 
Moreover, an important lemma ensures the covariance of the swapped distribution invariant, that is, a projection of $\wt{\bX}$ is also a generalized knockoff of $\bX^*$, presented in Supplementary material S.3.

\begin{remark}
    For $p < n < 2p$ , we can no longer compute the generalized knockoffs $\wt\bX$ by (\ref{GKnockoff2}), as there is no subspace of dimension $p$ orthogonal to $\bX^*.$ 
    Following \cite{BC2015}, 
    we can create row-augmented data and further use the Stab-GKnock, as long as an accurate estimate of $\sigma^2$ can be obtained.
\end{remark}

\vskip4mm
\noindent 
{\bf Step 2:  Construct generalized knockoff statistics.}
Once have generalized knockoff features, we need to construct generalized knockoff statistic $\bm{W}=(W_1,\dots, W_p)^{\T} \in \mathbb{R}^p$ as the testing statistic, obeying the sufficient property and the antisymmetry property mentioned in Section \ref{section2.1}. The type of statistic is not unique. A widely used choice is the LSM statistic, which is the point of the tuning parameter on Lasso regression path at which the feature first enters the model \citep{BC2015,BC2019,LSK2022}. 
However, the knockoff methods based on the LSM statistic may suffer power loss in real applications. 
{\cite{DGSZ2023} pointed out this problem, but did not provide a specific solution.}  

\begin{figure}[htbp]
    \begin{center}
    \includegraphics[width=1\textwidth,height=0.715267\textwidth]{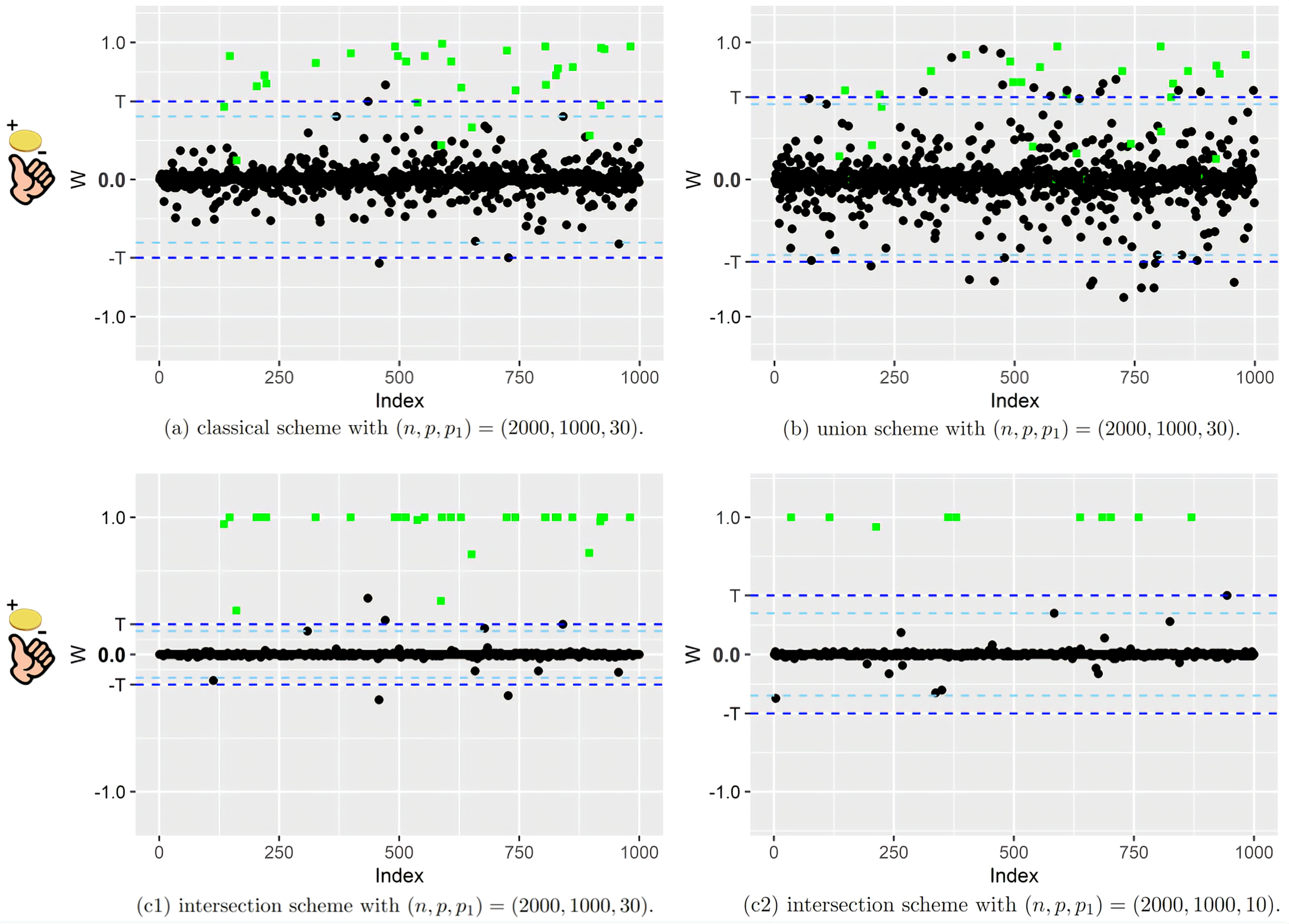}
    \end{center}
    \vspace{-5mm}
    \caption{{Scatter plot of $W_j$'s based on classical, union and intersection scheme with green squares denote true signals and black dots denote nulls, respectively. 
    The blue baseline denotes the threshold $T$ under $q=0.1$. 
    }
    }
    \label{scatter plot}
 \end{figure}

In this paper, we operate the subsampling strategy to enhance the selection stability, 
specifically using the projected spline estimator in Section \ref{section2.2} to obtain the selection probability as the variable importance score, and then construct the associated SPD statistic. 
However, a main concern about subsampling is that it will increase the variability of the selection result and face a power loss compared to other methods based on full data, as shown in Figure \ref{scatter plot}. 
To remedy this issue, we adopt an intersection subsampling-selection strategy similar to \cite{YFL2022,YKL2023}. 
Let $I \subset \{1,\dots,n\}$ denote the corresponding subsample indices with size $\lfloor n/2 \rfloor$, we obtain two projected spline estimators of the augment regression coefficient vector $\wt{\bbeta}=(\bbeta^{\T},\bm{0}_p^{\T})^{\T} \in \mathbb{R}^{2p}$ by (\ref{2.5}) using $([\bX^*,\wt{\bX}],\bm{Y}^*)(I)$ and $([\bX^*,\wt{\bX}],\bm{Y}^*)(I^c)$, respectively,
\begin{align}
    \label{augment estimator1}
    \wh{\bm{b}}(I)&=
    \arg \min_{\wt{\bm{\beta}} \in \mathbb{R}^{2p}}
    \left\{ \frac{1}{|I|} \|\bm{Y}^*(I)-[\bX^*,\wt{\bX}](I)\wt{\bbeta}\|^{2}+ 2\lambda \|\wt{\bbeta} \|_{1}\right\},\\
    \label{augment estimator2}
    \wh{\bm{b}}(I^c)&=
    \arg \min_{\wt{\bm{\beta}} \in \mathbb{R}^{2p}}
    \left\{ \frac{1}{|I^c|} \|\bm{Y}^*(I^c)-[\bX^*,\wt{\bX}](I^c)\wt{\bbeta}\|^{2}+ 2\lambda \|\wt{\bbeta} \|_{1}\right\},
\end{align}
{where $\lambda>0$ is a tuning parameter.} 
Then we get two estimates of $\mathcal{S}$ based on the subsample indices set $I$ and $I^c$ as
\begin{align}
    \label{classical strategy}
    \wh{\mathcal{S}}^{(1)}(I)&=\{j \in \{1,\dots,2p\}: \wh{b}_j(I) \neq 0\},\\
    \notag\label{indices2}
    \wh{\mathcal{S}}^{(2)}(I^c)&=\{j \in \{1,\dots,2p\}: \wh{b}_j(I^c) \neq 0\}.
\end{align}
We further adopt a simple intersection strategy  
to obtain the selected set $\wh{\mathcal{S}}(I)$ as follows
\begin{equation}
    \label{intersection strategy}
    \wh{\mathcal{S}}(I)=\wh{\mathcal{S}}^{(1)}(I) \cap \wh{\mathcal{S}}^{(2)}(I^c).
\end{equation}
Thus, the probability of being in the selected set $\wh{\mathcal{S}}(I)$ is
\begin{equation}
\label{SP}
    \wh{\Pi}_j= \mathbb{P}(j \in \wh{\mathcal{S}}(I)), \qquad j \in \{1,\dots,2p\},
\end{equation}
where $\mathbb{P}$ is taken concerning the randomness of subsampling. Further, we can define the generalized knockoff statistic $W_j$ for the transformed feature ${X}^*_j$ as {the SPD statistic}
\begin{equation}
\label{SPD}
    W_j= \wh{\Pi}_j - \wh{\Pi}_{j+p}, \qquad j \in \{1,\dots,p\}.
\end{equation}
As the selection probability $\wh{\Pi}_j$ is unknown, it can be estimated accurately by the empirical selection probability.
Specifically, we repeat the above subsampling and projected spline estimation procedure $L$ times, each time for two subsamples $I_l$ and $I_l^c$ for $l=1,\dots, L,$ and record 
the selected set as $\wh{\mathcal{S}}(I_l)=\wh{\mathcal{S}}^{(1)}(I_l) \cap \wh{\mathcal{S}}^{(2)}(I_l^c).$ Hence, we obtain the empirical selection probability for each variable $X_j$ based on $\{\wh{\mathcal{S}}(I_l)\}_{l=1}^L$, i.e.,
\begin{equation}
\label{ESP}
    \wt{\Pi}_j = \frac{1}{L} \sum_{l=1}^{L} 
    \bm{1}(j \in \wh{\mathcal{S}}(I_l)), \qquad j \in \{1,\dots,2p\}.
\end{equation}

\begin{remark}
   $\wh{\Pi}_j$ and $\wh{\Pi}_{j+p}$ can be regarded as the importance scores for the transformed feature $X^*_j$ and the generalized knockoff feature $\wt{X}_j$. 
   Noting that all generalized knockoff features are noises, a large positive $W_j$ indicates $X^*_j$ may be a true signal for $Y^*$, whereas a null $X^*_j$ induces $W_j$ is close to 0 and equally likely to be positive or negative \citep{BC2015,CFJL2018}. 
   The same can be said of the associated original feature $X_j$ for the response $Y$. 
   As pointed out in the literature, choosing $L = 100$ is sufficient to estimate the selection probability (\ref{SP}) accurately by (\ref{ESP}), we illustrate this point in Sections \ref{simu} and \ref{real}.
\end{remark}

{
Note that the proposed intersection selection strategy will help stabilize the selection result and boost the power. 
To illustrate this point, we consider a motivating simulation example. 
Generate each row of the design matrix $\bX$ independently from $N_p(\bm{0},\bSigma)$, where $\bSigma_{ij} = 0.6$ for $i\neq j$, $\bSigma_{ij} = 1$ for $i=j$. 
The sample size and the dimension are $(n,p)=(2000,1000)$.
We randomly set $p_1$ entries of the true regression parameter $\bm \beta$ to be nonzero with $p_1=10$ and $30$. These nonzero entries take values $\pm 1.2$ randomly.
We set the nonparametric smooth function $g(U) = \sin(2 \pi U)$, and the univariate $\{ U_i \}$ is i.i.d. drawn from the uniform distribution on $[0,1]$. 
In addition, the dimension $K$ of the B-spline basis is chosen by BIC criterion, and the tuning parameter $\lambda$ in (\ref{augment estimator1}) and (\ref{augment estimator2}) is chosen by 10-fold cross-validation.
Figure {\ref{scatter plot}} is the scatter plot of SPD statistic $W_j$ based on the classical selection strategy (\ref{classical strategy}) in \cite{MB2010}, the union selection strategy, and our intersection selection strategy (\ref{intersection strategy}). 
In all strategies, we can see that most $W_j$'s (green square) of the true signals are significantly positive, yet the $W_j$'s (black dot) of the nulls roughly symmetry about 0. 
Figures {\ref{scatter plot}}a and b show that, the union strategy and the classical strategy both cause $W_j$'s of nulls to severely inflate, which makes it difficult to distinguish the true signals from the nulls and hence results in power loss.
Conversely, Figures {\ref{scatter plot}}c1 and c2 depict that the intersection strategy sufficiently shrinks the statistics $W_j$'s of the nulls, which helps us identify the true signals and chooses a better threshold $T$. 
Moreover, the proposed intersection strategy also performs well when the signals become more sparse, illustrated by Figure {\ref{scatter plot}}c2.
In Section \ref{asy1}, we will further substantiate these points in Lemma \ref{lemma1} and Figure \ref{histogram}.
}

\begin{figure}[htbp]
    \begin{center} \includegraphics[width=1\textwidth,height=0.4774\textwidth]{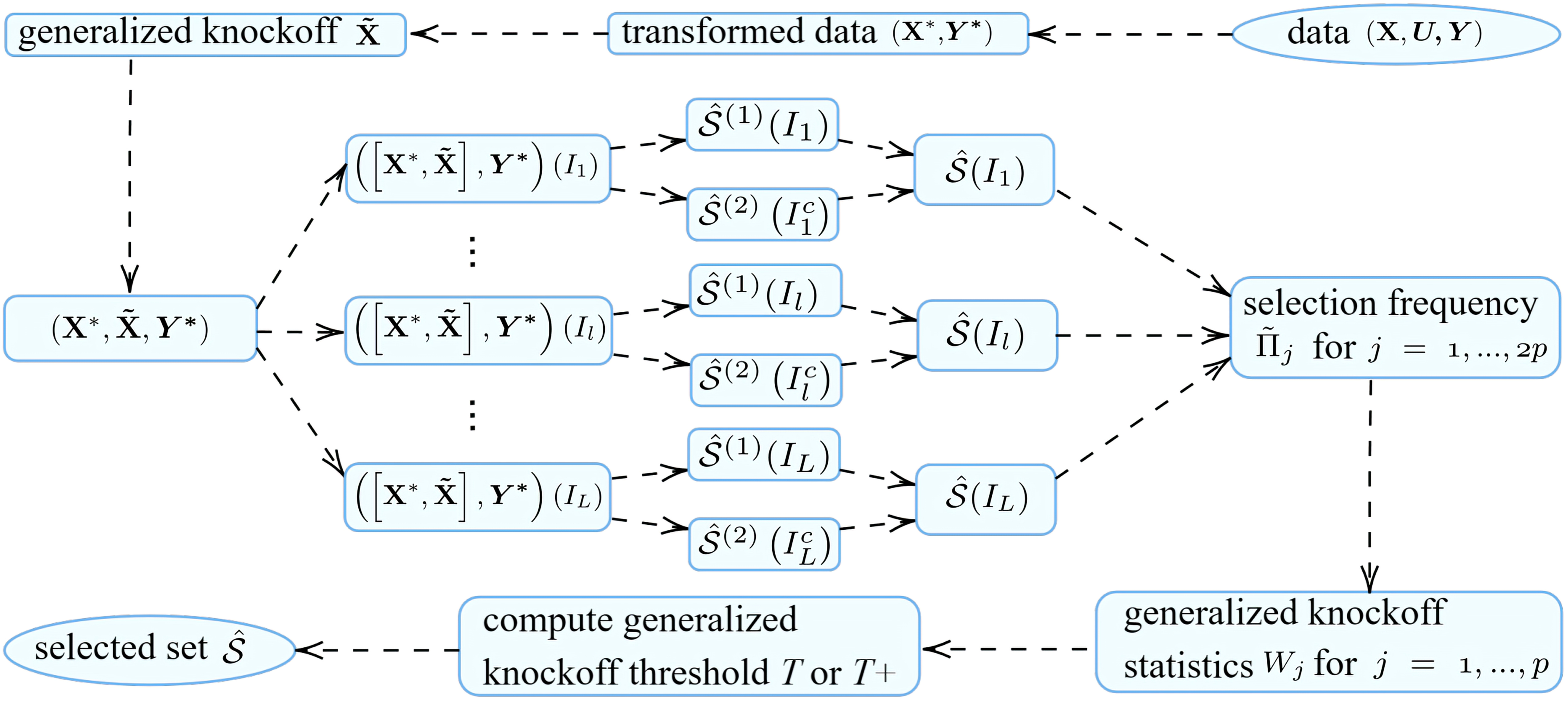}  
    \end{center}
    \caption{The workflow of the Stab-GKnock procedure.
    \label{workflow}}
 \end{figure}

\vskip4mm
\noindent 
{\bf Step 3: Select data-dependent threshold.} 
In our Stab-GKnock procedure, the final step is to choose a data-dependent threshold value $T$ via (\ref{threshold1}) or (\ref{threshold2}) mentioned in Section \ref{section2.1}. The active set $\mathcal{S}$ is estimated by
\begin{equation}
\label{Shat}
    \wh{\mathcal{S}}=\{j \in \{1,\dots,p\}: W_j \ge T\}.
\end{equation}
The workflow of the proposed Stab-GKnock procedure is presented in Figure \ref{workflow}.

\subsection{Stab-GKnock algorithm}
\label{sec Stab-GKnock algorithm}

In what follows, the Stab-GKnock procedure is summarized in Algorithm \ref{algo1}.

\begin{algorithm}[htbp]
	\caption{Stab-GKnock algorithm}
	\label{algo1}
	\LinesNumbered 
	\KwIn{The original data $(\bX,\bm{U},\bm{Y}) \in    
        \mathbb{R}^{n\times p} \times\mathbb{R}^{n} \times\mathbb{R}^{n}$, $p<n/2$, 
        the nominal FDR level $q\in (0,1)$, 
        {the degree of polynomial splines $m$, 
        }
        and the number of subsampling replications $L$.}
	\KwOut{The set of selected variables $\wh{\mathcal{S}}$.}

        \textbf{Step 1:}
	  Transform original data $(\bX, \bm{U}, \bm{Y})$ to $(\bX^*, \bm{Y}^*)$ based on splines' approximation and projection technique; \\
	\textbf{Step 2:}
	Construct the generalized knockoff matrix $\wt{\bX}$ to augment the transformed design matrix $\bX^*$ by (\ref{GKnockoff2}); \\
        
	\For{$l\gets 1$ to $L$}{
         \textbf{Step 3:}  
         Draw without put-back to obtain a subsample indices set $I_l \subset \{1,\dots,n\}$ and its complement set $I_l^c$, where $|I_l|=\lfloor n/2\rfloor$;\\
         \textbf{Step 4:}
         Solve the projection spline estimators $\wh{\bm{b}}(I_l), \wh{\bm{b}}(I_l^c)$ by (\ref{augment estimator1}) and (\ref{augment estimator2});\\
         \textbf{Step 5:}
         Record the intersected selected set $\wh{\mathcal{S}}(I_l)=\wh{\mathcal{S}}^{(1)}(I_l) \cap \wh{\mathcal{S}}^{(2)}(I_l^c)$ by (\ref{intersection strategy}).
            }
         \textbf{Step 6:}
         Compute the empirical selection probability $\{\wt{\Pi}_j\}_{j \in \{1,\dots,2p\}}$ by (\ref{ESP}), and then compute the generalized knockoff statistics $\{W_j\}_{j \in \{1,\dots,p\}}$ by (\ref{SPD});\\
         \textbf{Step 7:}
         Calculate the threshold $T$ by (\ref{threshold1}) or (\ref{threshold2}) at the nomial FDR level $q$;\\
         \textbf{Step 8:}
         Obtain the selected variables indices set $\wh{\mathcal{S}}=\left\{j \in \{1,\dots,p\}: W_j \ge T\right\}$.
        
\end{algorithm}

\subsection{Theoretical results}
\label{asy1}
\begin{figure}[H]
    \begin{center}
    \includegraphics[width=1\textwidth,height=0.348485\textwidth]{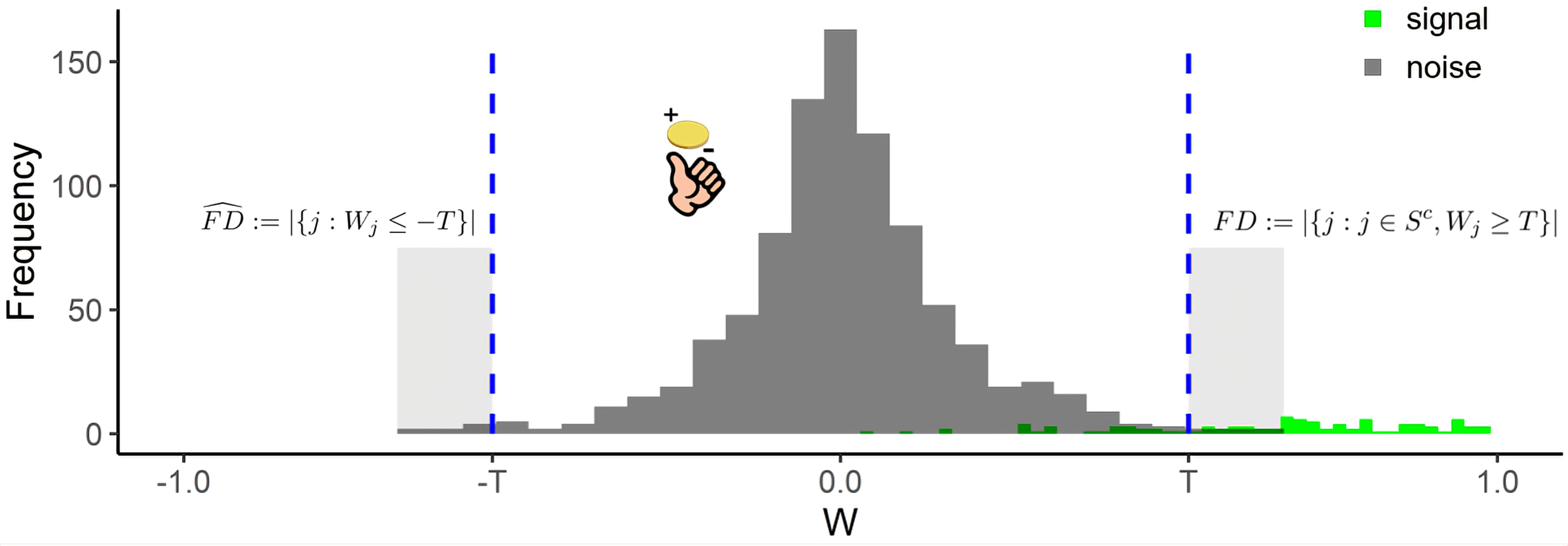}
    \end{center}
    \vspace{-10mm}
    \caption{
    {
    The sampling distribution histogram of generalized knockoff statistic $W_j$. Set $n=2000$, $p=1000$, $p_1=100$, $A=0.6$, $\rho = 0.5$, $\{\bm{x}_i\} \stackrel{\mathrm{i.i.d.}}{\sim} N_p(\bm{0}, \bSigma)$, where $\bSigma_{ij} = \rho $ for $i\neq j$, $\bSigma_{ij} = 1$ for $i = j$. The nonparametric function $g(U) = \sin(2 \pi U)$, where $\{ U_i \} \stackrel{\mathrm{i.i.d.}}{\sim} U(0,1).$
    The blue baseline denotes the threshold $T$ under $q=0.1$.
    Variables with $W_j>T$ are selected, and an overestimate of false discoveries (\text{FD}) can be obtained owing to the sign-flip property.
    }
    }
    \label{histogram}
 \end{figure}

In this subsection, we build theoretical guarantees for the Stab-GKnock procedure in terms of FDR and power.
We first give the following lemma to show the sign-flip property (\ref{sign-flip}) of the statistics (\ref{SPD}), which is also illustrated in Figure \ref{histogram}.

\begin{lemma}[Sign-flip property for the nulls] \label{lemma1} 
Let $\{\epsilon_1,\dots,\epsilon_p\}$ be a set of independent random variables, such that $\epsilon_j=1 ~\text{if}~ j\in \mathcal{S}, ~\text{and}~ \epsilon_j = \pm 1$ with equal probability $1/2$ if $j \in \mathcal{S}^c$. Then,
    \begin{equation*}
        (W_1,\dots,W_p) \overset{d}{=} (\epsilon_1 \cdot W_1,\dots, \epsilon_p \cdot W_p).
    \end{equation*}
\end{lemma}
\noindent
Lemma \ref{lemma1} is the key of our proposed Stab-GKnock framework, which gives an ``overestimate'' of FDP.
To see this, for any $t \geq 0$, we have
\begin{equation*}
    \frac{ {|\left\{ j:W_{j}\geq t, j \in \mathcal{S}^c \right\}|} }{|\left\{j:W_{j}\geq t \right\}| \vee 1} \approx
    \frac{ {|\left\{ j:W_{j}\leq -t, j \in \mathcal{S}^c \right\}|} }{|\left\{j:W_{j}\geq t \right\}| \vee 1}\leq
    \frac{ {|\left\{ j:W_{j}\leq -t \right\}|} }{|\left\{j:W_{j}\geq t \right\}| \vee 1} := \widehat{\mathrm{FDP}}(t).
\end{equation*}
\noindent

{
Consider a similar simulation example in Section \ref{sec Stab-GKnock}, the sampling distribution histogram of generalized knockoff statistic $W_j$ is presented in Figure \ref{histogram}, where the green squares and black dots denote true signals and nulls, respectively. 
We can see that most $W_j$'s of the true signals are significantly positive, yet the $W_j$'s of the nulls roughly symmetry about 0.
Owing to Lemma \ref{lemma1}, we can select variables with $W_j \geq T$, and conservatively estimate FDP using the left tail of the distribution.
}

The proof of Lemma \ref{lemma1} is presented in Supplementary Material S.4. 
The following theorem indicates that the Stab-GKnock procedure can control FDR at the nominal level $q \in (0,1)$ for any finite sample size $n$.

\begin{theorem}[FDR control of Stab-GKnock] \label{th3}
For any $q \in (0,1)$, choose the threshold $T > 0$ by (\ref{threshold1}). Then the Stab-GKnock procedure which retains the set
\begin{equation*}
    \wh{\mathcal{S}}=\{j \in \{1,\dots,p\}: W_j \ge T\}
\end{equation*}
controls the modified FDR defined as
\begin{equation*}
    {\rm mFDR}(\widehat{\mathcal{S}}) = \mathbb{E}\left[\frac{|\widehat{\mathcal{S}} \cap \mathcal{S}^c|}{|\widehat{\mathcal{S}}| + 1/q}\right] \leq q
\end{equation*}
for any finite sample size $n$. If the threshold $T > 0$ is chosen by (\ref{threshold2}), then the Stab-GKnock controls the usual FDR
\begin{equation*}
    {\rm FDR}(\widehat{\mathcal{S}}) = \mathbb{E}\left[\frac{|\widehat{\mathcal{S}} \cap \mathcal{S}^c|}{|\widehat{\mathcal{S}}| \vee 1} \right] \leq q.
\end{equation*}
\end{theorem}
\noindent
The proof of Theorem \ref{th3} counts on the result of Lemma \ref{lemma1} and the original proof in \cite{BC2015}. 
Next, we are inquisitive about the other side of the coin, that is, the power guarantee of our proposed Stab-GKnock procedure. In order to establish the theoretical results, we need some basic regularity conditions.

\begin{Condition}[Minimal signal condition] \label{C1}
    There exists some slowly diverging sequence $\kappa_n \to \infty$, such that $$\min_{j\in \mathcal{S}} |\bbeta_j| \geq \kappa_n \sqrt{\log{(2p)}/n}, \quad~\text{as}~n \to \infty.$$
\end{Condition}

\begin{Condition}[Minimal eigenvalue condition] \label{C2} 
    There exist a constant $C_1 > 0$, such that
    \begin{equation*}
        \lambda_{\min}
        \left(\frac{1}{n}
        \left[\mathbf{X}^*_{\mathcal{S}}, \widetilde{\mathbf{X}}_{\mathcal{S}}\right]^{\rm{T}}
        \left[\mathbf{X}^*_{\mathcal{S}}, \widetilde{\mathbf{X}}_{\mathcal{S}}\right]\right)
        \geq C_1.
    \end{equation*}
\end{Condition}

\begin{Condition}[Mutual incoherence condition] \label{C3}
    There exists a constant $\gamma_{\scriptscriptstyle I} \in (0,1]$, which may depend on the subsampling index $I$, such that  
    \begin{equation*}
    \hspace{-1mm}
        \max_{j \in \mathcal{S}^c}
        \left\|  \left[\mathbf{X}^*(I), \widetilde{\mathbf{X}}(I)\right]_{j}^{\rm{T}}\!
        \left[\mathbf{X}^*_{\mathcal{S}}(I), \widetilde{\mathbf{X}}_{\mathcal{S}}(I)\right]\!
        \left(\!\left[\mathbf{X}^*_{\mathcal{S}}(I), \widetilde{\mathbf{X}}_{\mathcal{S}}(I)\right]^{\rm{T}}\!
        \left[\mathbf{X}^*_{\mathcal{S}}(I), \widetilde{\mathbf{X}}_{\mathcal{S}}(I)\right]\!\right)^{-1}  \right\|_{2} \!\leq\! 1-\gamma_{\scriptscriptstyle I} .
    \end{equation*}
\end{Condition}

Conditions \ref{C1}--\ref{C3} are crucial for establishing the asymptotic power result for Stab-GKnock.
Condition \ref{C1} is a signal strength condition, which ensures that the projected spline estimator (\ref{augment estimator1}) does not miss too many true signals. Condition \ref{C1} is easily satisfied in high-dimensional regression, see \cite{Zhang2009}. 
Condition \ref{C2} is known as the minimal eigenvalue condition, which states that the Gram matrix of the active set on the augment design matrix is invertible.
Condition \ref{C3} is known as the mutual incoherence condition, which indicates that the correlation between the true signals and nulls should not be too strong.
Conditions \ref{C2} and \ref{C3} are common technique conditions for Lasso regression \citep{ZY2006,Wainwright2019}, 
which further ensure the variable selection consistency for the projected spline estimator (\ref{augment estimator1}). 
\begin{remark}
    As the original design matrix $\bX$ and the projection matrix $\bW$ are observable, we impose Conditions \ref{C1}--\ref{C3} on the transformed design matrix $\bX^*=\bW \bX$ instead of $\bX$. Similar treatments can be found in \cite{MH2015} and \cite{LSK2022}. 
\end{remark}

\begin{theorem}[Power of Stab-GKnock]
\label{th4}
    Let $\lambda \asymp \sqrt{\log (2p)/n}$, and suppose that regularity conditions \ref{C1}--\ref{C3} hold. Then, the selected set $\wh{\mathcal{S}}$ obtained by Stab-GKnock satisfies
\begin{equation*}
    {\rm Power}(\widehat{\mathcal{S}}) = \mathbb{E}\left[\frac{|\widehat{\mathcal{S}} \cap \mathcal{S}|}{|\mathcal{S}| \vee 1} \right] \rightarrow 1, \quad~\text{as}~n \rightarrow \infty.
\end{equation*}
\end{theorem}

Theorem \ref{th4} indicates that the Stab-GKnock attains an asymptotic full power, that is, it can identify all important features asymptotically as $n \to \infty$. The proof of Theorem \ref{th4} is presented in Supplementary Material S.6, which is enlighted by \cite{FDLL2020}. 
Essentially, the Stab-GKnock procedure relies on the selection results of the projected spline estimator (\ref{augment estimator1}). Thus, we need Lasso-type regression (\ref{augment estimator1}) to achieve the variable selection consistency, i.e., it can identify all true signal variables. See S.1 in Supplementary Material for more details.

\section{High-dimensional setting: SPLS-Stab-GKnock}
\label{HD-setting}
The Stab-GKnock procedure in Section \ref{metho} demands $n>2p$ and is not applicable for high-dimensional setting. In this section, we propose a two-stage procedure based on data-splitting technique. 
Specifically, we split the full data into two parts. 
In the first screening step, we implement a newly proposed joint screening method, called Sparse-PLS, to reduce the dimension $p$ to a suitable dimension $\wh{p}_1$ using the first part data of size $n_1$. 
In the second selection step, we further apply Stab-GKnock to select the variables on the screened variables set using the second part data of size $n_2$, where $n=n_1+n_2$. 
We first introduce the Sparse-PLS screening method in Section \ref{sec SPLS}, and then summarize the two-stage algorithm in Section \ref{sec SPLS-Stab-GKnock algorithm}.

\subsection{Joint screening procedure in PLM: Sparse-PLS}
\label{sec SPLS}

Note that our two-stage procedure for high-dimensional setting is a natural extension of the low-dimensional Stab-GKnock procedure as long as the screening step correctly captures all relevant features.
Hence, we desire the sure screening property in \cite{FL2008} 
to be attained.
Most existing screening methods rely on the marginal effects of features on the response, such as SIS \citep{FL2008} and RRCS \citep{LPZZ2012}.
There are two main concerns about marginal screening methods. First, despite screening based on marginal effects having computational efficiency, they are often unreliable in practice since they ignore the joint effects of candidate features. Second, the feature with significant joint effect but weak marginal effect is likely to be wrongly left out by marginal screening methods. We illustrate these points in Sections \ref{simu} and \ref{real}.

Here, we propose a joint screening strategy for high-dimensional PLM via the sparsity-constrained projected least squares estimation (Sparse-PLS, SPLS). Considering the high-dimensional partially linear model (\ref{model2}) in Section \ref{section2.2}, we obtain the following projected least squares objective function of $\bbeta$ by splines’ approximation and projection technique
\begin{equation}
\label{PLS}
    \mathcal{L}(\bbeta)=
    \frac{1}{2n} \|\bm{Y}-\bX \bbeta-\bZ \btheta\|^{2} = 
    \frac{1}{2n} \|\bW(\bm{Y}-\bX \bbeta)\|^{2} = 
    \frac{1}{2n} \|\bm{Y}^*-\bX^*\bbeta\|^{2}.
\end{equation}
Suppose $\bbeta$ is sparse, i.e., the true model size $p_1 \leq k$ for some user-specified sparsity $0<k<p$. The proposed Sparse-PLS estimator can be defined as
\begin{equation}
\label{Sparse-PLS}
    \wh{\bbeta}(k)=
    \arg \min_{\bm{\beta} \in \mathbb{R}^p} 
    {\mathcal{L}(\bbeta)},
    \quad ~{\rm s.t.}~ \|\bbeta \|_{0} \le k.
\end{equation}
Then the screened set of Sparse-PLS is obtained
\begin{equation}
\label{screened set}
    \wh{\mathcal{S}}_1=
    \{j \in \{1,\dots,p\}: \wh{\beta}_j(k) \neq 0\}.
\end{equation}

The sparsity constraint $\|\bbeta \|_{0} \le k$ in (\ref{Sparse-PLS}) specifies the number of features allowed in the model, that is, the Sparse-PLS procedure just estimates some of the coefficients while presets the others to 0, which makes Sparse-PLS suitable for feature screening. 
Note that estimation (\ref{Sparse-PLS}) is carried out on the full model, $\wh{\bbeta}(k)$ can be viewed as a screener which naturally accounts for the joint effects among features and hence goes beyond marginal utilities. 
Essentially, the Sparse-PLS proposes a sparsity-constraint estimator by adopting a $\mathcal{L}_0$-regularization technique, which has similarities with the SMLE in \cite{XC2014} and the constrained Danzig selector (CDS) in \cite{KZL2016}. 




On the other side of the coin, the proposed Sparse-PLS procedure can also be viewed as a high-dimensional best subset selection with subset size $k$ \citep{BKM1967,Miller2002}, 
thereby the cardinality constraint makes problem (\ref{Sparse-PLS}) become an NP-hard problem \citep{Natarajan1995}. 
In this article, we solve the screener $\wh{\bbeta}(k)$ and implement Sparse-PLS efficiently by using a modern optimization method, specifically mixed integer optimization (MIO).
It can obtain a near-optimal solution efficiently for the nonconvex optimization problem (\ref{Sparse-PLS}), theoretically shown in \cite{BKM2016}. 
Finally, 
we prove the Sparse-PLS computed by MIO enjoys the sure screening property under some regularity conditions.
We start with some additional regularity conditions. 

\begin{Condition}[NP-dimensionality condition]
\label{C4}
    Let $\log(p)=O(n^\kappa)$ for some $0 \leq \kappa <1$.
\end{Condition}  

\begin{Condition}[Minimal signal condition]
\label{C5}
    There exist some nonnegative constants $\omega_1, \omega_2, \tau_1$ and $\tau_2$ such that
    $\min_{j \in \mathcal{S}} |\beta_j| \geq \omega_1 n^{-\tau_1}$, and $p_1 \leq k \leq \omega_2 n^{-\tau_2}$. 
\end{Condition}

\begin{Condition}[UUP condition]
\label{C6}
    There exist constants $c_1>0$ and $\delta_1>0$ such that for sufficiently large $n$, 
    $$\lambda_{\min}(n^{-1} {\bX^*_{\mathcal{A}}}^{\T}\bX^*_{\mathcal{A}}) \geq c_1$$
    for $\mathcal{A} \in \bS^{2k}_{+}$ and $\bbeta_\mathcal{A} \in \{\bbeta_\mathcal{A}:\|\bbeta_\mathcal{A}-\bbeta^*_\mathcal{A}\| \leq \delta_1 \}$, 
    where $\bS^{k}_{+} \coloneqq \{\mathcal{A}: \mathcal{S} \subset \mathcal{A}; |\mathcal{A}|\leq k\}$
    and $\bS^{k}_{-} \coloneqq \{\mathcal{A}: \mathcal{S} \not\subset \mathcal{A}; |\mathcal{A}| \leq k\}$
    denote the collections of the over-fitted models and the under-fitted models, respectively.
\end{Condition}

\begin{Condition}[Dependence condition]
\label{C7}
     There exist constants $c_2, c_3>0$ such that $|X^*_{ij}|\leq c_2$ and 
    $$\max_{1 \leqslant j \leqslant p} \max_{1 \leqslant i \leqslant n} \left\{\frac{{X^*_{ij}}^{2}}{\sum_{i=1}^{n} {X^*_{ij}}^{2} {\sigma^*_i}^2}\right\} \leq c_{3} \cdot n^{-1},$$
    when n is sufficiently large, where 
    ${\sigma^*_i}^2=\var({Y}_i^*)$.
\end{Condition}

Condition \ref{C4} imposes an assumption that $p$ can diverge up to an exponential rate with $n$, which means the dimension $p$ can be greatly larger than the sample size $n$. 
Condition \ref{C5} permits the coefficients of true signal variables to degenerate slowly as $n$ diverges, which is widely used in the literature of screening methods \citep[see, e.g., ][]{FL2008,LPZZ2012,LKLL2022}
It also places a weak restriction on the sparsity $k$ to make sure screening possible.
Condition \ref{C6} restricts the pairwise correlations between the columns of $\bX^*$ in consideration, which is equivalent to the UUP condition given in \cite{CT2007}. 
This condition is mild and commonly used in high-dimensional methods, like DS \citep{CT2007}, SIS-DS \citep{FL2008}, FR \citep{Wang2009}, SMLE \citep{XC2014}, CDS \citep{KZL2016}, GFR \citep{CFLL2018} and CKF \citep{SXZ2020}. 
Condition \ref{C7} is also a restriction on the transformed design matrix $\bX^*$, which holds naturally so long as ${\sigma^*_i}^2$ does not degenerate too fast, noted by \cite{XC2014}. 
Under Conditions \ref{C4}--\ref{C7}, the following Theorem \ref{th5} states the sure screening property.

\begin{theorem}[Sure screening property]
\label{th5}
Assume regularity conditions \ref{C4}--\ref{C7} hold with $\tau_1+\tau_2<\frac{(1-\kappa)}{2}$, and $\wh{\mathcal{S}}_1$ is the MIO computed screened set of size $k$ from the Sparse-PLS procedure, we have
\begin{equation*}
    \mathbb{P}(\mathcal{S} \subset \wh{\mathcal{S}}_1) \to 1, \quad
    ~\text{as}~n \rightarrow \infty.
\end{equation*}
\end{theorem}

Theorem \ref{th5} ensures that the subset selected by Sparse-PLS would not miss any true signal variable with probability tending to one. 
The proof of Theorem \ref{th5} is given in Supplementary Material S.7.

\begin{remark}
     The sparsity $k$ controls the threshold between signal and null features, 
     thus the choice of $k$ is a key point in screening procedures. Standard hard-threshold choices often set $k=c\lfloor n/\log(n) \rfloor$ for some $c>0$ \citep{FL2008,LPZZ2012}, 
     it can also be selected by a data-driven strategy in \cite{GRZL2023}. 
     In simulation studies, we find the proposed Sparse-PLS has robust performance for a wide choice of $k$ compared with the marginal methods, shown in Table \ref{tab_screening1}.

\end{remark}


\subsection{High-dimensional SPLS-Stab-GKnock algorithm}
\label{sec SPLS-Stab-GKnock algorithm}
In this subsection, the two-stage SPLS-Stab-GKnock procedure is extended based on data-splitting technique, which is briefly introduced as follows
\begin{itemize}
    \item 
    \begin{sloppypar}   
    Randomly split the data into two groups 
    $(\bX^{(1)},\bm{U}^{(1)},\bm{Y}^{(1)}) \in \mathbb{R}^{n_1\times p} \times\mathbb{R}^{n_1} \times\mathbb{R}^{n_1}$ and 
    $(\bX^{(2)},\bm{U}^{(2)},\bm{Y}^{(2)}) \in \mathbb{R}^{n_2\times p} \times\mathbb{R}^{n_2} \times\mathbb{R}^{n_2}$, where $n_1+n_2=n$.
    \end{sloppypar}
    
    \item \textsc{Screening step}: Apply Sparse-PLS on $(\bX^{(1)},\bm{U}^{(1)},\bm{Y}^{(1)})$ and obtain the screened set $\wh{\mathcal{S}}_1$ which reduces the dimension $p$ to a suitable dimension $\wh{p}_1<n_2/2$. 

    \item \textsc{Selection step}: Apply Stab-GKnock to further select variables on the screened variables set $\wh{\mathcal{S}}_1$ using $(\bX^{(2)},\bm{U}^{(2)},\bm{Y}^{(2)})$.
    
\end{itemize}

We summarize the SPLS-Stab-GKnock procedure in Algorithm \ref{algo2}.

\begin{remark}
Along with Condition \ref{C5}, we also require $k \le n_2/2$ in the screening step, which is not only necessary for establishing the sure screening property but ensures a suitable dimension $\wh{p}_1 \coloneqq \|\wh{\bbeta}(k) \|_{0} \leq n_2/2$ for the subsequent Stab-GKnock selection step.
\end{remark}

\begin{algorithm}[htbp]
	\caption{SPLS-Stab-GKnock algorithm}
	\label{algo2}
	\KwIn{The original data $(\bX,\bm{U},\bm{Y})\in 
        \mathbb{R}^{n\times p} \times\mathbb{R}^{n} \times\mathbb{R}^{n}$, the data-splitting indices sets $\mathcal{N}_1 \in \{1,\dots,n\},~\mathcal{N}_2=\{1,\dots,n\} \setminus \mathcal{N}_1$, $|\mathcal{N}_1|=n_1,~|\mathcal{N}_2|=n_2=n-n_1,~p_1 \leq n_2/2$, 
        the pre-screened size $k$ with $p_1 \leq k \leq n_2/2$, 
        the nominal FDR level $q\in (0,1)$, 
        {the degree of polynomial splines $m$, 
        }
        and the number of subsampling replications $L$.}
	\KwOut{The set of selected variables $\wh{\mathcal{S}}$.}

        \textbf{Step 1:}
	  Split the data into two groups 
        $(\bX^{(1)},\bm{U}^{(1)},\bm{Y}^{(1)})\in \mathbb{R}^{n_1\times p} \times\mathbb{R}^{n_1} \times\mathbb{R}^{n_1}$ and $(\bX^{(2)},\bm{U}^{(2)},\bm{Y}^{(2)})\in \mathbb{R}^{n_2\times p} \times\mathbb{R}^{n_2} \times\mathbb{R}^{n_2}$; \\

        \textbf{Screening Step: Sparse-PLS}\\
        \quad\textbf{Step 2:}
        Run the Sparse-PLS screening procedure on $(\bX^{(1)},\bm{U}^{(1)},\bm{Y}^{(1)})$, compute the Sparse-PLS screener $\wh{\bbeta}(k)$ by (\ref{Sparse-PLS});\\
        
        \quad\textbf{Step 3:}
        Select the screened set $\wh{\mathcal{S}}_1$ by (\ref{screened set});\\
        
        \textbf{Selection Step: Stab-GKnock}\\
        \quad\textbf{Step 4:}
	  Transform the restricted data 
        $(\bX^{(2)}_{\wh{\mathcal{S}}_1}, \bm{U}^{(2)}, \bm{Y}^{(2)})$ to 
        $({\bX^{(2)}_{\wh{\mathcal{S}}_1}}^*, \bm{Y}^*)$ based on splines' approximation and projection technique; \\
	\quad\textbf{Step 5:}
	Construct the generalized knockoff matrix $\wt{\bX}^{(2)}_{\wh{\mathcal{S}}_1}$ to augment ${\bX^{(2)}_{\wh{\mathcal{S}}_1}}^*$ by (\ref{GKnockoff2}); \\
        
	\For{$l\gets 1$ to $L$}{
         \textbf{Step 6:}  
         Draw without put-back to obtain a subsample indices set  $I_l \subset \mathcal{N}_2$ and its complement set $I_l^c=\mathcal{N}_2 \setminus I_l$, where $|I_l|=\lfloor n_2/2\rfloor$;\\
         \textbf{Step 7:}
         Solve the projection spline estimators $\wh{\bm{b}}(I_l)$ and $\wh{\bm{b}}(I_l^c)$ based on data $\left(\left[{\bX^{(2)}_{\wh{\mathcal{S}}_1}}^*, \wt{\bX}^{(2)}_{\wh{\mathcal{S}}_1}\right], {\bm{Y}^{(2)}}^*\right)$ by (\ref{augment estimator1}) and (\ref{augment estimator2}), respectively;\\
         \textbf{Step 8:}
         Record the selected set $\wh{\mathcal{S}}(I_l)=\wh{\mathcal{S}}^{(1)}(I_l) \cap \wh{\mathcal{S}}^{(2)}(I_l^c)$ by (\ref{intersection strategy}).
         }
         \quad\textbf{Step 9:}
         Compute the empirical selection probability $\{\wt{\Pi}_j\}$ by (\ref{ESP}), and then compute the generalized knockoff statistics $\{W_j\}_{j \in \wh{\mathcal{S}}_1}$ by (\ref{SPD});\\
         \quad\textbf{Step 10:}
         Calculate the threshold $T$ by (\ref{threshold1}) or (\ref{threshold2}) at the nominal FDR level $q$;\\
         \quad\textbf{Step 11:}
         Obtain the selected variables indices set $\wh{\mathcal{S}}=\{j \in \wh{\mathcal{S}}_1: W_j \ge T\}$.       
\end{algorithm}

\begin{remark}
{
    For high-dimensional settings, the two-stage procedure using data-splitting technique is commonly considered in knockoff-based literature \citep{BC2019, FDLL2020}.
    There are two main concerns about these methods. 
    On the one hand, the screening accuracy determines the performance of the subsequent selection.
    Despite enjoying the sure screening property asymptotically, the marginal screening method used in the first step causes the two-stage procedure to suffer power loss in real applications, mentioned in \cite{SXZ2020} and \cite{LKLL2022}. 
    On the other hand, data splitting increases the variability of the selection result and may cause the loss of statistical power in practice.
    For future research, we suggest that an in-depth analysis of the data-splitting strategy could potentially enhance the performance even further, such as the unequal subsample size strategy or overlapping splitting strategy.
    In addition, the ``data recycling'' idea proposed in \cite{BC2019} can be borrowed to improve the power.
}
\end{remark}

\subsection{Theoretical results}
\label{asy2}
In this subsection, we establish theoretical guarantees for the SPLS-Stab-GKnock procedure in terms of FDR and power. Denote $\mathcal{E}=\{\mathcal{S} \subset \wh{\mathcal{S}}_1\}$ the event that the screening step possesses the sure screening property.

\begin{theorem}[FDR control of SPLS-Stab-GKnock]
\label{th6}
Under regularity conditions \ref{C4}--\ref{C7}, for any $q \in (0,1)$, choose the threshold $T > 0$ by (\ref{threshold1}), the SPLS-Stab-GKnock procedure satisfies 
\begin{equation*}
    {\rm FDR}(\widehat{\mathcal{S}}) = \mathbb{E}\left[
    \frac{|\widehat{\mathcal{S}} \cap \mathcal{S}^c|}{|\widehat{\mathcal{S}}| \vee 1} 
    \right] \leq q
\end{equation*}
with the probability tending to one as $n \to \infty$. Furthermore, conditioning on the sure screening event $\mathcal{E}=\{\mathcal{S} \subset \wh{\mathcal{S}}_1\}$ and
select the threshold $T$ by (\ref{threshold2}), then we obtain a finite-sample FDR control guarantee
\begin{equation*}
    {\rm FDR}(\widehat{\mathcal{S}}) = \mathbb{E}\left[\frac{|\widehat{\mathcal{S}} \cap \mathcal{S}^c|}{|\widehat{\mathcal{S}}| \vee 1} 
    \arrowvert \mathcal{E} \right] \leq q.
\end{equation*}
\end{theorem}
Theorem \ref{th6} states that conditioning on event $\mathcal{E}$, the SPLS-Stab-GKnock procedure achieves FDR control at the nominal level $q \in (0,1)$.

\begin{theorem}[Power of SPLS-Stab-GKnock]
\label{th7}
Under regularity conditions \ref{C4}--\ref{C7} and conditions in Theorem \ref{th5}. Then, conditional on the sure screening event $\mathcal{E}=\{\mathcal{S} \subset \wh{\mathcal{S}}_1\}$ and select the threshold $T$ by (\ref{threshold2}), the SPLS-Stab-GKnock procedure satisfies
\begin{equation*}
    {\rm Power}(\widehat{\mathcal{S}}) = \mathbb{E}\left[
    \frac{|\widehat{\mathcal{S}} \cap \mathcal{S}|}{|\mathcal{S}| \vee 1}  \arrowvert \mathcal{E} \right] \rightarrow 1, \quad~\text{as}~n \rightarrow \infty.
\end{equation*}
\end{theorem}
Theorem \ref{th7} indicates that conditioning on event $\mathcal{E}$, the SPLS-Stab-GKnock procedure attains an asymptotic full power as $n$ diverges to infinity.

\section{Simulation studies}
\label{simu}   

In this section, we conduct numerical simulations to evaluate the finite-sample performance of the proposed Stab-GKnock, Sparse-PLS screening and SPLS-Stab-GKnock procedure. In Section \ref{simu-low}, we consider the finite sample performance of Stab-GKnock for $n \geq 2p$ case. In Section \ref{simu-screening}, we evaluate the screening performance of SPLS. In Section \ref{simu-high}, we assess the performance of SPLS-Stab-GKnock for $p>n$ case. 
{In all cases, we set $L = 100$, and the tuning parameter $\lambda$ is selected by cross-validation.}

\subsection{Low-dimensional performance}
\label{simu-low}

In this subsection, we evaluate the empirical performance of Stab-GKnock in low-dimensional cases. Specifically, we consider the partially linear model (\ref{model1}) with $n \geq 2p$. We draw each row of the design matrix $\bX$ independently from

\begin{itemize}
\setlength{\abovedisplayskip}{3pt}
\setlength{\belowdisplayskip}{3pt}	
	\item Case 1: A centered multivariate Gaussian distribution $N_p(\bm{0},\bSigma)$, where $\bSigma=(\rho^{|i-j|})_{1 \leq i,j \leq p}$ for correlations $\rho=0.2$ and $0.5$;
	\item Case 2: A centered multivariate $t$ distribution $t_{p,3}(\bm{0},\bSigma)$ with degrees of freedom 3, where $\bSigma=(\rho^{|i-j|})_{1 \leq i,j \leq p}$ for correlations $\rho=0.2$ and $0.5$.
\end{itemize}

We randomly set $p_1$ entries of the true regression parameter $\bm \beta$ to be nonzero. These nonzero entries take values $\pm A$ randomly with $A=0.2,0.4,0.6,0.8$ and $1.0$. 
We set the nonparametric smooth function $g(U) = \sin(2 \pi U)$, and the univariate $\{ U_i\}$ is i.i.d. drawn from the uniform distribution on $[0,1]$. 
{We set the spline order $m=3$, and set the number of internal knots $K^*=n^{1/9}$, which is the theoretically optimal order.}
The errors $\{ \epsilon_i \}$ are i.i.d. copies from $N(0,1)$. 
{It is worth noting that the Stab-GKnock procedure demonstrates robust performance for a wide choice of error variance ${\sigma}^{2}$.
Relevant simulations are not pursued here due to space limitations.}


We set the desired FDR level as $q = 0.1$, and choose the tuning parameter $\lambda$ in (\ref{2.5}) by 10-fold cross-validation {using R package \texttt{glmnet}}. We compare the following five methods for different settings based on 200 replications.

\begin{itemize}
	\setlength{\abovedisplayskip}{3pt}
	\setlength{\belowdisplayskip}{3pt}	
	\item Stab-GKnock: The proposed procedure in this paper implemented via Algorithm \ref{algo1} with the threshold selected by (\ref{threshold1}).
	\item Stab-GKnock+: The proposed procedure in this paper implemented via Algorithm \ref{algo1} with the threshold selected by (\ref{threshold2}).
        \item B-H: The BH procedure applied to p-values from univariate regression in \cite{BH1995}, {which is implemented using function ``univglms'' in R package \texttt{Rfast}.} 
        \item Knock-LSM+: The fixed-X knockoff procedure with LSM knockoff statistic in \cite{BC2015}, {which is implemented using function ``create.fixed'' in R package \texttt{knockoff}. We generate the knockoff features with  \textit{equicorrelated} construction to choose $\bm{s}$.}
	\item m-Knock+: The model-X knockoff procedure with Lasso coefficient difference {(LCD)} statistic in \cite{CFJL2018}, {which is implemented using function ``create.guassian'' in R package \texttt{knockoff}. We use the second-order approximation in \cite{CFJL2018} to generate the knockoff features with \textit{approximate semidefinite program} construction to choose $\bm{s}$.}  
	
\end{itemize}
\begin{remark}
	Note that the B-H procedure and the fixed-X knockoff procedure are proposed for linear models, we first apply the projection technique to convert model (\ref{model1}) to a linear model, and then apply two procedures on the transformed model.
\end{remark}

\begin{figure}[htbp]
  \centering
  \includegraphics[width=1\textwidth,height=0.71434\textwidth]{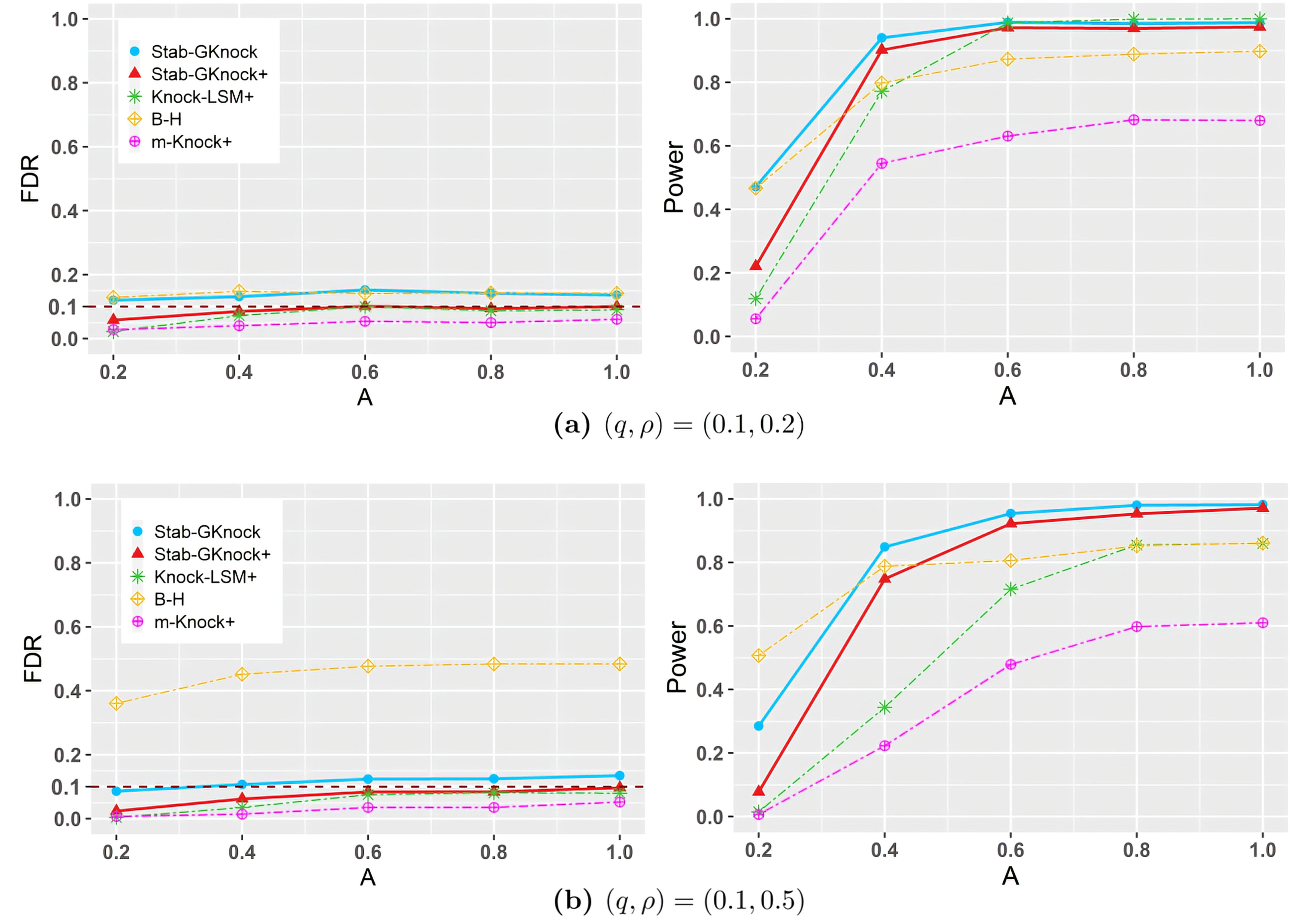}
  \vspace{-10mm} \caption{{
  Empirical FDRs and powers for Case 1 with $n=300$, $p=150$, $p_1=20$, $q=0.1$, $\rho \in \{0.2,0.5\}$ and $A \in \{0.2,0.4,0.6,0.8,1.0\}$. 
  The red dashed lines indicate the desired FDR level.
  } 
  }
  \label{low_dimension_1}
  
\end{figure}

\begin{figure}[htbp]
  \centering
  \includegraphics[width=1\textwidth,height=0.72581\textwidth]{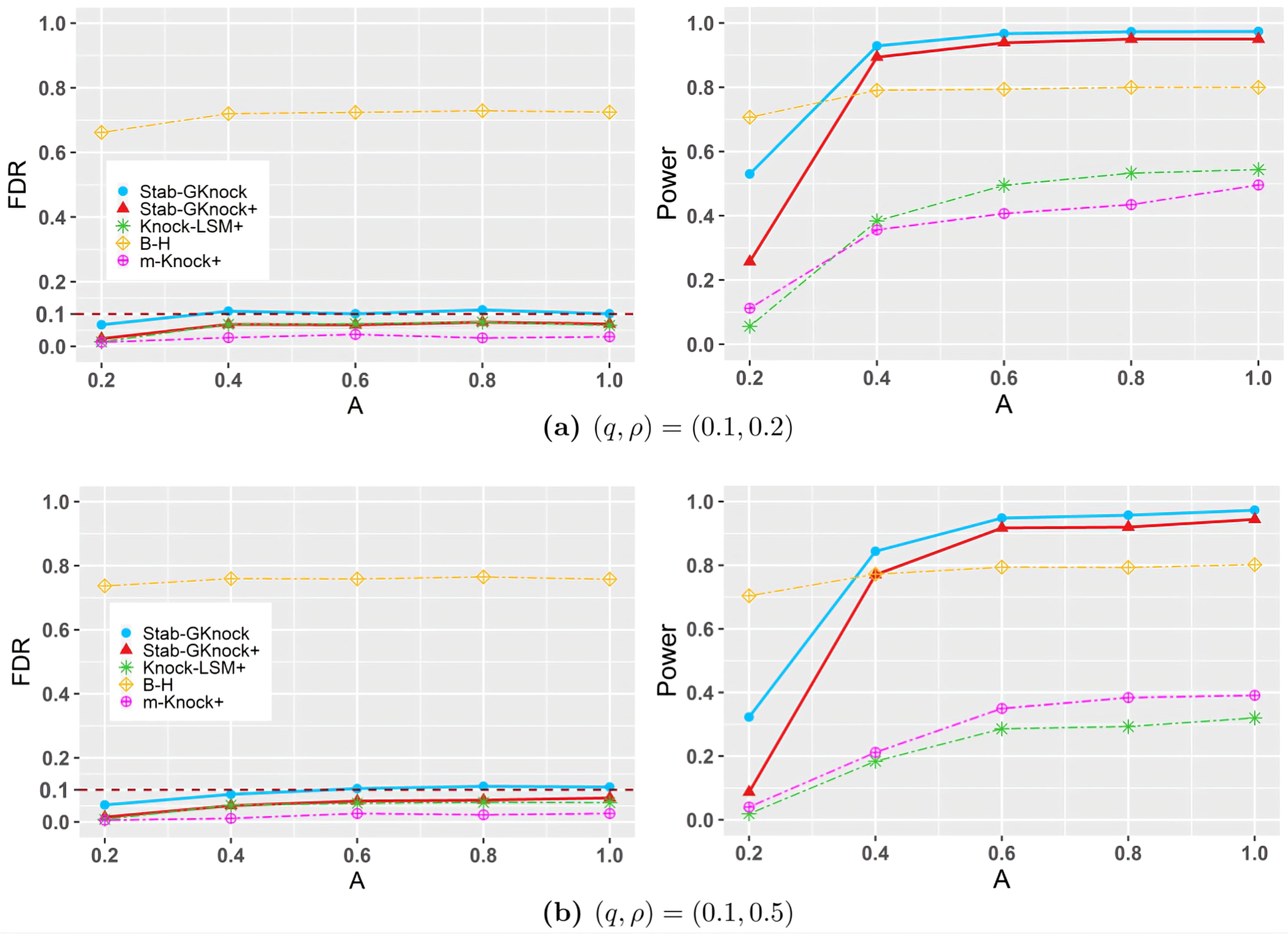}
  \vspace{-10mm} \caption{Empirical FDRs and powers for Case 2 with $n=300$, $p=150$, $p_1=20$, $q=0.1$, $\rho \in \{0.2,0.5 \}$ and $A \in \{0.2,0.4,0.6,0.8,1.0\}$.
  The red dashed lines indicate the desired FDR level.
  }
  \label{low_dimension_2}
  
\end{figure}

\begin{figure}[htbp]
  \centering
  \includegraphics[width=1\textwidth,height=0.73347\textwidth]{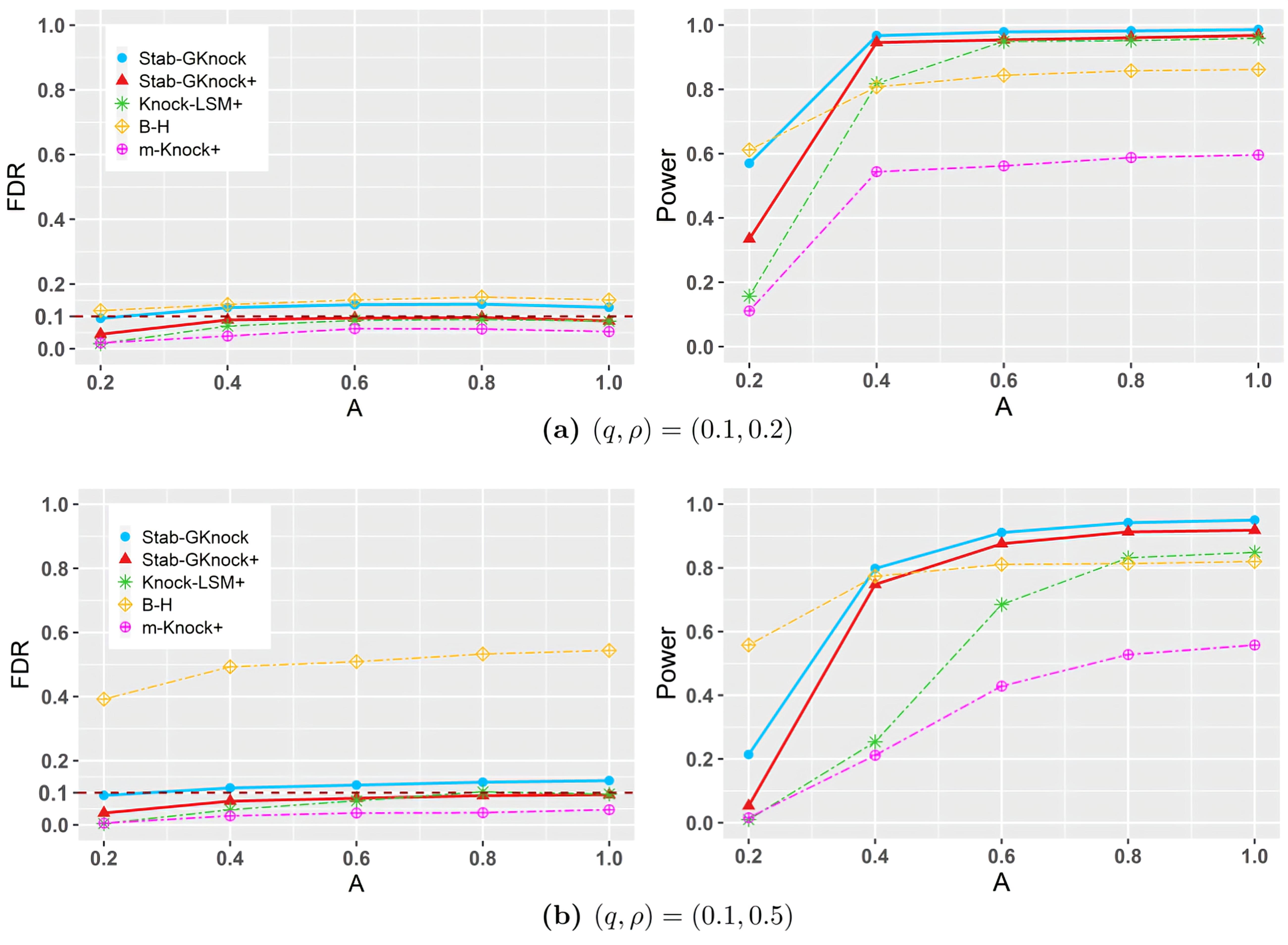}
  \vspace{-10mm} \caption{Empirical FDRs and powers for Case 1 with $n=400$, $p=200$, $p_1=20$, $q=0.1$, $\rho \in \{0.2,0.5 \}$ and $A \in \{0.2,0.4,0.6,0.8,1.0\}$.
  The red dashed lines indicate the desired FDR level.}
  \label{low_dimension_3}

\end{figure}

Figures \ref{low_dimension_1} to \ref{low_dimension_3} show the simulation results for $n \geq 2p$ cases. We can observe that the proposed method demonstrates favorable results in most settings, and presents significant improvement in power compared to other procedures. More specifically, we have the following findings.

(1) Figure \ref{low_dimension_1} reports the low-dimensional simulation results when $\bX$ is Gaussian design. We find that the power of all methods rises with increasing signal strengths $A$, yet the Stab-GKnock procedures always yield basically the highest power. Meanwhile, the proposed methods are not sensitive when the correlation $\rho$ increases from $0.2$ to $0.5$, while the powers of Knock-LSM+ and m-Knock+ slightly decline. In addition, although B-H method maintains high power as $\rho$ increases, it can not control FDR at target level $q=0.1$ anymore.

(2) Figure \ref{low_dimension_2} examines whether our proposed methods still work when $\bX$ is non-Gaussian design. We find that all the methods successfully control FDR except for the B-H procedure, and the powers of our proposed Stab-GKnock methods still tend to one. On the contrary, the power of Knock-LSM+ exhibits a noticeable decrease compared to the Gaussian scenario above, as it is tailored for Gaussian design. The model-X knockoff method is relatively robust for the design matrix. 

(3) Figure \ref{low_dimension_3} examines whether our proposed methods perform well when the signals become more sparse. We find that our proposed Stab-GKnock methods still have the highest power as the signal sparsity $p_1 / p$ descends. We attribute it to the intersection strategy of our SPD statistics $W_j$ defined in (\ref{SPD}), which is visually illustrated in Figure \ref{scatter plot}.

In summary, FDR is well controlled for our proposed methods,  even when the design matrix is not normally distributed or the signals are sparse. Apart from its robustness, Stab-GKnock also holds remarkably higher power compared to the main competitors across different scenarios.

\subsection{Screening performance}
\label{simu-screening}

In this subsection, we assess the performance of the proposed Sparse-PLS procedure for high-dimensional screening. Specifically, we set $p=700$, $n=200$ and $A=0.6$ with design matrix $\bX$ generated by Case 1 or Case 2. 
We measure the screening performance by calculating: (1) FDR, the empirical average false discovery rate after screening; (2) PRR, the averaged proportion of signals that are retained after screening; (3) SSR, the proportion of times that all signals are retained after screening; (4) MMS, the minimum model size to include all signals. We compare the following five methods under the same setup based on 200 replications.

\begin{itemize}
	\setlength{\abovedisplayskip}{3pt}
	\setlength{\belowdisplayskip}{3pt}	
	\item SPLS: The proposed procedure in this paper which selects the screened set by (\ref{screened set}).
	\item SIS: The sure independence screening procedure based on Pearson's $\rho$ correlation coefficient in \cite{FL2008}. 
        \item RRCS: The robust rank correlation screening procedure based on Kendall's $\tau$ correlation coefficient in \cite{LPZZ2012}. 
        \item PFR: The profiled forward regression algorithm in \cite{LWT2012}. 
	\item SPLasso: The sequential profile Lasso method in \cite{LLT2017}. 
	
\end{itemize}
\begin{remark}
 We apply SIS and RRCS procedures after employing our spline approximation and projection technique in Section \ref{section2.2}, 
 and the profiled techniques used in PFR and SPLasso are also replaced by spline approximation. 
 Moreover, SMLE in \cite{XC2014} is a likelihood-based method that is not suitable for partially linear models, and CDS in \cite{KZL2016} focuses on the convergence rates under weak signal strength assumption, not feature screening, hence we do not compare with these methods considering joint effects mentioned above.
\end{remark}

\begin{table}[htbp]
    \centering
    \caption{Screening accuracy for Case 1 with {predefined sparsity} $k=40$.}
    \vspace{3mm}
	\begin{tabular*}{\textwidth}{@{\extracolsep{\fill}}ccccccc}
		\hline\hline
		&  & SPLS & SIS & RRCS & PFR & SPLasso \\ 
            \hline\hline
            \multicolumn{1}{c}{\multirow{3}{*}{$p_1=20$}} & FDR & 0.502 & 0.641 & 0.667 & 0.502 & 0.506 \\
		\multicolumn{1}{c}{} & PRR & 0.995 & 0.717 & 0.666 & 0.996 & 0.987 \\
		\multicolumn{1}{c}{} & SSR & 0.960 & 0.000 & 0.000 & 0.950 & 0.810 \\ \hline
		\multicolumn{1}{c}{\multirow{3}{*}{$p_1=30$}} & FDR & 0.296 & 0.618 & 0.636 & 0.370 & 0.421 \\
		\multicolumn{1}{c}{} & PRR & 0.939 & 0.510 & 0.486 & 0.839 & 0.772 \\
		\multicolumn{1}{c}{} & SSR & 0.610 & 0.000 & 0.000 & 0.330 & 0.110 \\ \hline
		\multicolumn{1}{c}{\multirow{3}{*}{$p_1=40$}} & FDR & 0.361 & 0.589 & 0.605 & 0.500 & 0.511 \\
		\multicolumn{1}{c}{} & PRR & 0.639 & 0.411 & 0.396 & 0.500 & 0.489 \\
		\multicolumn{1}{c}{} & SSR & 0.030 & 0.000 & 0.000 & 0.000 & 0.000 \\ 
            \hline\hline
	\end{tabular*}
        \label{tab_screening1}
\end{table}

\begin{table}[htbp]
    \centering
    \caption{Quantiles of MMS for Case 2 with {the numbers of the relevant features} $p_1=20$.}
    \vspace{3mm}
    \begin{tabular*}{\textwidth}{@{\extracolsep{\fill}}l *{5}{>{\raggedleft\arraybackslash}p{1.1cm}}}
        \hline\hline
        \multicolumn{1}{c}{Method/MMS}& 
        \multicolumn{1}{r}{5\%} & \multicolumn{1}{r}{25\%} & \multicolumn{1}{r}{50\%} & \multicolumn{1}{r}{75\%} & \multicolumn{1}{r}{95\%} \\
        \hline \hline
        SPLS & 20 & 21 & 22 & 22 & 25 \\
        SIS & 402 & 554 & 633 & 678 & 698 \\ 
        RRCS & 181 & 291 & 435 & 564 & 668 \\
        PFR & 21 & 24 & 26 & 29 & 38 \\
        SPLasso & 22 & 26 & 30 & 41 & 665\\ \hline
        \hline
    \end{tabular*}    
    \label{tab_screening2}
\end{table}

Tables \ref{tab_screening1} and  \ref{tab_screening2} summarize the simulation results for screening. We can find that the proposed Sparse-PLS procedure achieves a promising screening accuracy compared to the marginal methods. It can identify the majority of signals after screening to a desirable model size. More specifically, we have the following findings.

(1) Table \ref{tab_screening1} reports the screening accuracy with fixed model size $k = 40 \approx n/\log(n)$. Noting that a smaller FDR with a larger PRR and SSR suggests a more accurate screening method, we find that the proposed SPLS method performs remarkably well as $p_1$ varies. Its highest SSR guarantees a high power of the subsequent selection analysis, which is in line with its theoretical property in Theorem \ref{th5}. In contrast, the other SIS-based marginal methods are likely to be affected by the correlation among features. Unlike SPLS, which can jointly assess the significance of covariates, they tend to leave out some relevant features.

(2) Table \ref{tab_screening2} shows $5\%$, $25\%$, $50\%$, $75\%$, and $95\%$ quantiles of the minimum model size to include all signals when $\bX$ is non-Gaussian design, where a smaller quantile indicates a more effective screening approach. We find that the proposed SPLS method can detect the signals with the smallest model size. It allows us to sharply decrease the number of features in the first stage of screening. In contrast, SIS and RRCS demand a larger model to cover all signals, which illustrates that the feature with significant joint effect but weak marginal effect is likely to be wrongly left out by these marginal screening methods. 
By introducing Kendall's $\tau$ correlation, RRCS achieves better performance in non-Gaussian design but still does not screen out the majority of nulls. 
Moreover, PFR performs quite well since its strategy helps to incorporate some feature joint effects in the screening process compared to SIS, yet faces a high computational cost and is still inferior to SPLS.

To summarize, the numerical results are in line with the sure screening property of the SPLS procedure, which allows a high power in the second stage of Algorithm \ref{algo2} for subsequent selection. It is worth noting that the screening size $k$ is set as the hard threshold in this paper. To make the $k$ statistically interpretable, treating it as a tuning parameter within the model can also be an interesting topic for future research.


\subsection{High-dimensional performance}
\label{simu-high}

In this subsection, we conduct the two-stage SPLS-Stab-GKnock procedure for high-dimensional cases. Specifically, we set $(n,p,p_1)=(500,1500,20)$ for the partially linear model (\ref{model1}) with design matrix $\bX$ generated by Case 1 or Case 2. As illustrated in Algorithm \ref{algo2}, we randomly divide the full data into two parts and use $n_1=250$ samples to conduct the SPLS procedure for the screening step. After reducing to $k=100$ variables, we use the remaining samples to compare the Stab-GKnock procedure with Knock-LSM+, B-H and m-Knock+ for the selection step. All the following results are based on 200 replications, and the target FDR level is also set as $q=0.1$.

\begin{figure}[t]
  \centering
  \includegraphics[width=1\textwidth,height=0.72296\textwidth]{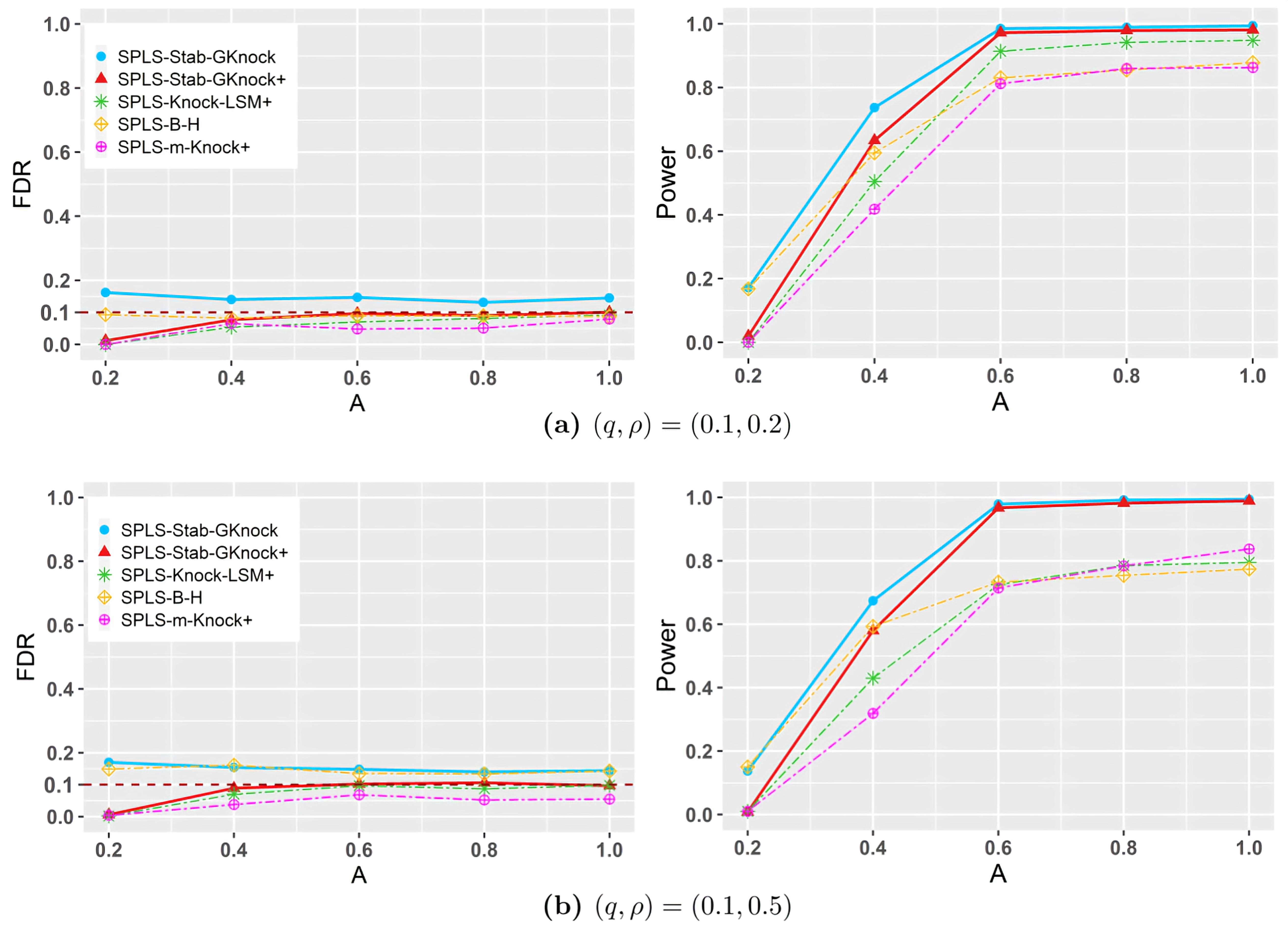}
  \vspace{-10mm} \caption{Empirical FDRs and powers for Case 1 with $n=500$, $p=1500$, $p_1=20$, $q=0.1$, $\rho \in \{0.2,0.5 \}$ and $A \in \{0.2,0.4,0.6,0.8,1.0\}$.
  The red dashed lines indicate the desired FDR level.}
  \label{high_dimension_1}
  
\end{figure}

\begin{figure}[t]
  \centering
  \includegraphics[width=1\textwidth,height=0.72381\textwidth]{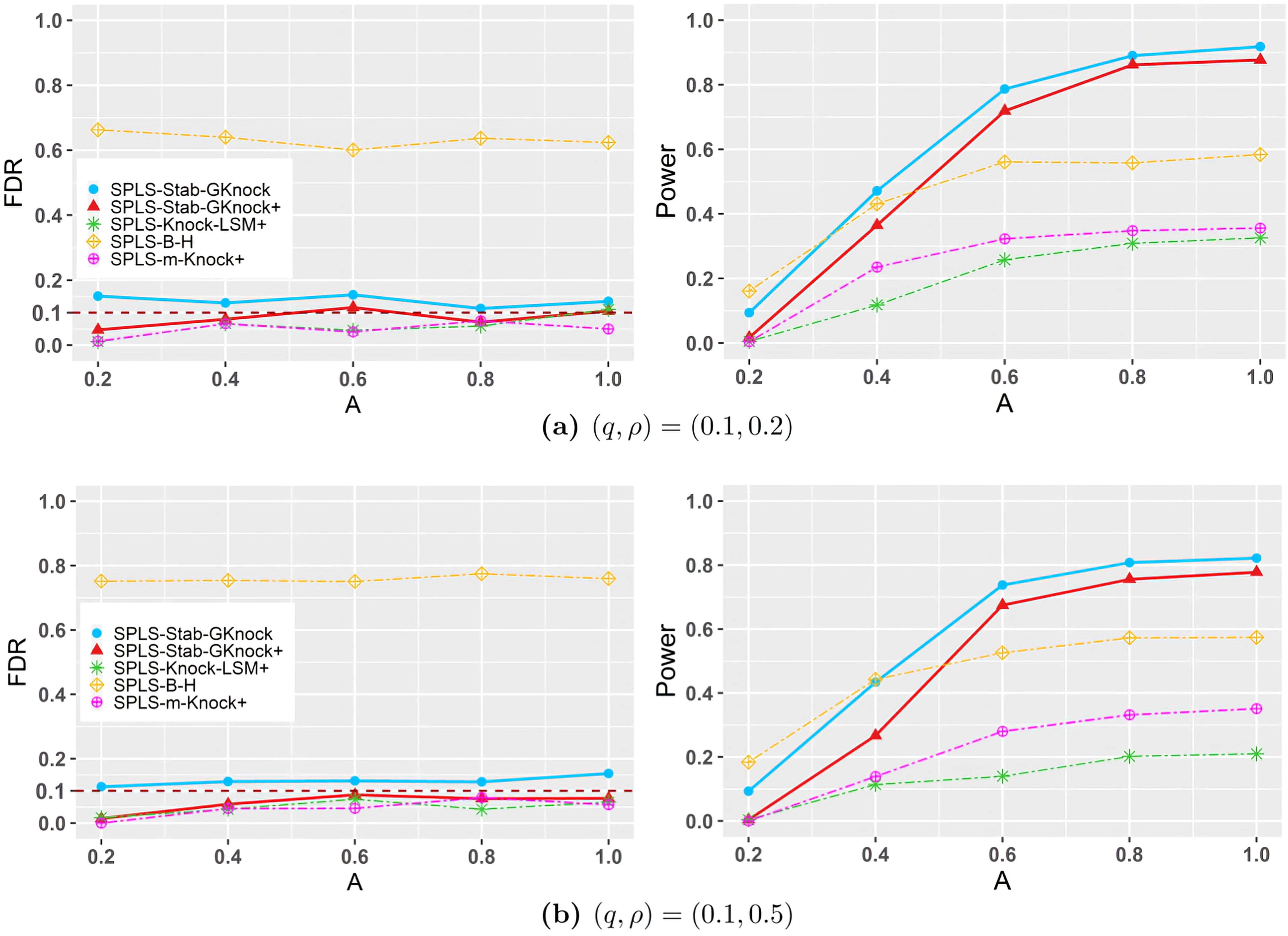}
  \vspace{-10mm} \caption{Empirical FDRs and powers for Case 2 with $n=500$, $p=1500$, $p_1=20$, $q=0.1$, $\rho \in \{0.2,0.5 \}$ and $A \in \{0.2,0.4,0.6,0.8,1.0\}$.
  The red dashed lines indicate the desired FDR level.
  }
  \label{high_dimension_2}
  
\end{figure}

Figures \ref{high_dimension_1} and \ref{high_dimension_2} show the simulation results for $p>n$ cases. Our two-stage SPLS-Stab-GKnock procedure performs well in terms of FDR and power. Specifically, we have the following findings.

(1) Figure \ref{high_dimension_1} reports the finite-sample simulation results when $\bX$ is Gaussian design. We find that the SPLS-Stab-GKnock procedure successfully controls FDR. Its power tends to one as the signal strength $A$ increases, which also confirms that SPLS will not lose many signals in the first screening stage. In comparison, SPLS-Knock-LSM+ and SPLS-B-H methods are sensitive when increasing the correlation from $0.2$ to $0.5$, and SPLS-m-Knock+ results in a significant power loss.

(2) Figure \ref{high_dimension_2} shows the finite-sample performances when $\bX$ is non-Gaussian design. We find that the proposed SPLS-Stab-GKnock still enjoys the highest power while controlling FDR at the target value. On the contrary, the SPLS-B-H method fails to control FDR in non-Gaussian scenarios. The SPLS-Knock-LSM+ and SPLS-m-Knock+ are too conservative to achieve a desirable high power.

In a nutshell, our two-stage procedure can perfectly handle high-dimensional cases with regard to both FDR and power.

\section{Real data analysis}
\label{real}
In this section, we illustrate the effectiveness of our proposed methods by an application to a breast cancer dataset, which has been analyzed in \cite{YLM2012} and \cite{CFLL2018}. 
As reported in \cite{WSW2014}, 
breast cancer is the most common cancer diagnosis in women across 140 countries and is the most frequent cause of cancer mortality in 101 countries. \cite{vant2002} 
collected a breast cancer dataset from 97 lymph node-negative breast cancer patients under 55 years old.
This dataset contains 97 rows and 24481 columns, each row contains 24481 gene expression levels and 7 clinical risk factors including age, tumour size, histological grade, angioinvasion, lymphocytic infiltration, estrogen receptor (ER) and progesterone receptor status for 97 patients. By removing the missing genes, we can obtain 24188 gene expressions. In this section, ER is regarded as the response supported by the study in \cite{KLGM1977}, 
and the patient's age is regarded as the univariate for nonparametric component. Both the response and covariates have been standardized with mean zero and variance one.

The goal is to identify genes that are related to the ER of breast cancer patients. We consider the following high-dimensional partially linear model
\begin{equation}
	\label{6.1}
	Y_i=\bm{X}_{i}^{\T}\bbeta + g(U_i) + \varepsilon_{i}, \qquad i=1,\dots,97,
\end{equation}
where $Y_i$ is the ER of the breast cancer patient, $\bm{X}_{i}$ is the \textit{p}-dimensional covariates vector consisted of 24188 genes expressions, $\bbeta=({\beta}_1,\dots,{\beta}_p)^{\T}$ is a \textit{p}-dimensional vector of unknown regression coefficients, $U_i$ is the patient's age.

To deal with this problem, we perform our proposed SPLS-Stab-GKnock in two stages as illustrated in Algorithm \ref{algo2}. Specifically, we randomly select $n_1=50$ samples to conduct the SPLS procedure for model (\ref{6.1}) and obtain $k=23$ candidate genes in the screening step. Then we use the remaining 47 samples to apply the Stab-GKnock procedure for final selection with target FDR level $q = 0.2$. 
We compare our proposed method with Lasso, SPLS-B-H, SPLS-Knock-LSM+ and SPLS-m-Knock+ based on 200 replications. 
{
We apply Lasso after employing spline approximation and projection technique in Section \ref{section2.2}.
}
\begin{table}[htbp]
    \centering
    \caption{{Sample mean and sample standard error (in parentheses) of model size for the breast cancer dataset study.}}
    \vspace{3mm}
    \begin{tabular*}{\textwidth}{p{13cm}c}
    \hline\hline
        Methods &  Model size \\ \hline\hline
        Lasso &  25.16(2.24)\\
        SPLS-Stab-GKnock & 10.11(0.71) \\
        SPLS-Stab-GKnock+ & 7.39(0.64) \\
        SPLS-B-H &  17.80(2.88)\\
        SPLS-Knock-LSM+ &  2.72(0.88) \\
        SPLS-m-Knock+ &  4.06(0.68)\\ \hline\hline
    \end{tabular*}
    \label{tab_real1}
\end{table}

Table \ref{tab_real1} briefly summarizes the sample mean and standard error (in parentheses) of the model sizes selected by each method. We find that our SPLS-Stab-GKnock obtains a moderate model size among all the methods. It neither selects too many genes like Lasso and SPLS-B-H nor is it as overly conservative as SPLS-m-Knock+ and SPLS-Knock-LSM+, which could lead to the potential loss of some relevant genes.

\begin{table}[H]
    \centering
    \caption{
    {
    Selected genes by Lasso, SPLS-Stab-GKnock+ and SPLS-m-Knock+ for the breast cancer dataset study.
    }
    }
    \vspace{3mm}
    \begin{tabular*}{\textwidth}{@{\extracolsep{\fill}}c*{3}{>{\centering\arraybackslash}p{0.225\textwidth}}}
    \hline\hline
    \multirow{2}{*}{Gene} & \multicolumn{3}{c}{Methods} \\ \cline{2-4} 
     & Lasso & SPLS-Stab-GKnock+ & SPLS-m-Knock+ \\ \hline\hline
     $X_{15835}$ & \checkmark & \checkmark & \checkmark \\
    $X_{10478}$ & \checkmark &  &  \\
    $X_{1690}$ & \checkmark & \checkmark &  \\
    $X_{13695}$ & \checkmark & \checkmark & \checkmark \\
    $X_{1279}$ & \checkmark & \checkmark & \checkmark \\
    $X_{8979}$ & \checkmark &  & \\
    $X_{20331}$ & \checkmark &  &  \\
    $X_{6912}$ & \checkmark & \checkmark &  \\
    $X_{12188}$ & \checkmark &  & \\
    $X_{3109}$ & \checkmark &  &  \\
    $X_{10177}$ & \checkmark & \checkmark &  \\
    \hline\hline
    \end{tabular*}
    \label{tab_real2}
\end{table}

Table \ref{tab_real2} presents genes selected by Lasso, SPLS-Stab-GKnock+ and SPLS-m-Knock+ more than 80 percent of repetitions. We find that all three methods select genes 15835, 13695 and 1279. The proposed SPLS-Stab-GKnock+ method also selects genes 1690, 13695, 6912 and 10177, and excludes several genes selected by Lasso, which partly echoes the results in \cite{CHZ2016} and \cite{LLT2017}. 

\begin{figure}[htbp]
  \centering
  \includegraphics[width=1\textwidth,height=0.5\textwidth]{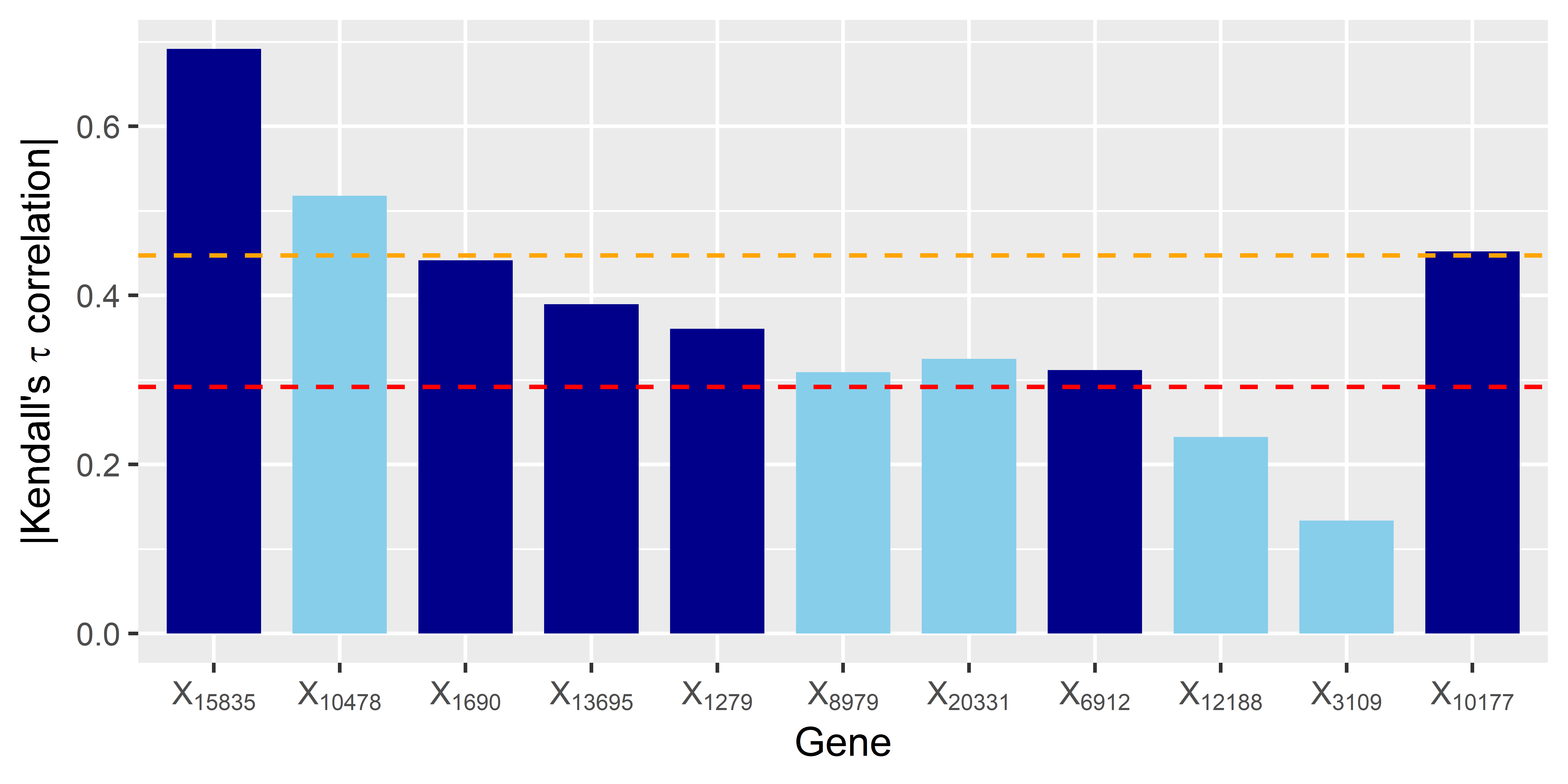}
  \vspace{-10mm} \caption{The absolute value of Kendall's $\tau$ correlation with regard to the response. Note that the {\color{blue} 
 blue} bars represent genes selected by both Lasso and SPLS-Stab-GKnock, and the {\color{skyblue} 
 skyblue} bars represent genes only selected by Lasso. The {\color{orange} orange} and {\color{red} red} dashed lines indicate the 23rd and 1000th highest absolute value of Kendall's $\tau$ correlation with regard to the response, respectively.}
 \label{barplot}
  
\end{figure}

\begin{figure}[htbp]
  \centering
  \includegraphics[width=1\textwidth,height=0.5\textwidth]{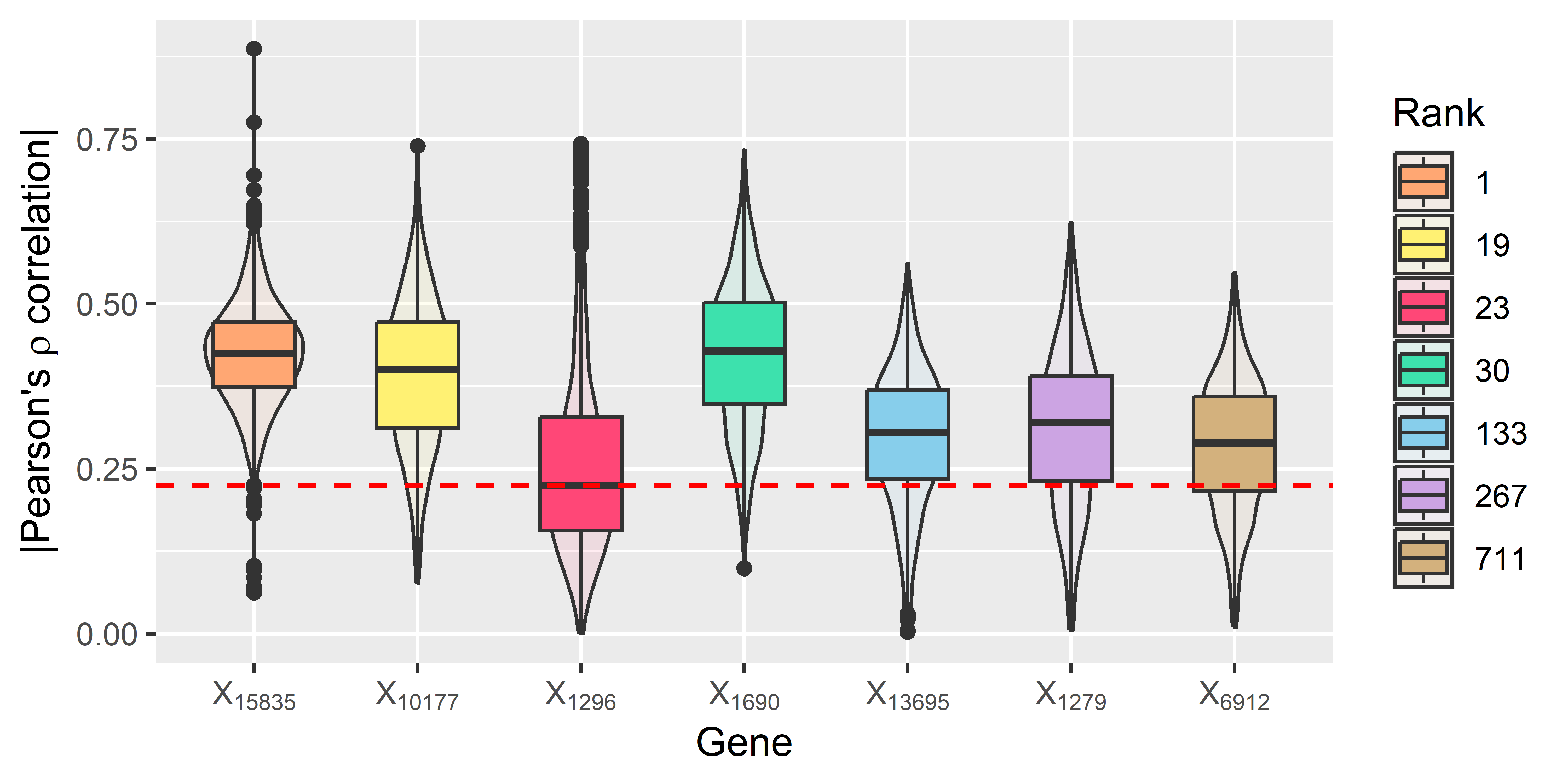}
  \vspace{-10mm} \caption{
  {
  The boxplot about the absolute value of Pearson's $\rho$ correlation with regard to the top 1000 genes (other than itself), genes on the horizontal axis are ranked by Kendall's $\tau$ correlation with the response.
  }
  }
  \label{boxplot}
  
\end{figure}

To confirm our conjecture that SPLS can jointly evaluate the significance of covariates in Section \ref{sec SPLS}, we further compare the marginal and joint effects of genes selected by SPLS-Stab-GKnock. 
Figure \ref{barplot} reports the marginal effects of genes on the response. 
{To ensure the subsequent selection step works, we have to obtain at most 23 genes in the screening step. We can see that among all 6 genes selected by SPLS-Stab-GKnock, only genes 15835 and 10177 are among the top 23 genes ranked by Kendall's $\tau$ correlation with the response ER. In other words, if we use RRCS for screening, two-thirds of genes selected by SPLS-Stab-GKnock will be left out.}
Similar results will be achieved when other marginal-based methods, like SIS, are employed for screening.

Although bearing relatively lower marginal effects on the response, Figure \ref{boxplot} demonstrates that genes selected by SPLS-Stab-GKnock exhibit stronger joint effects on other relevant genes. 
Specifically, we use Pearson's $\rho$ correlation between genes to evaluate their joint effects, and set the performance of gene 1296, the 23th gene ranked by Kendall's $\tau$ correlation with the response ER, as the baseline. 
Genes in the boxplot are arranged in descending order by their Kendall's $\tau$ correlation coefficients with regard to the response, as depicted in the legend for their rankings among all genes. 
From the boxplot, we find that genes 15835 and 10177 with stronger marginal effects (on the left side) are selected by both SPLS-Stab-GKnock and RRCS, while genes 1690, 13695, 1279 and 6912 with lower marginal effects but stronger joint effects (on the right side) are selected only by SPLS-Stab-GKnock. In other words, all the genes selected by SPLS-Stab-GKnock share higher joint effects on other associated genes compared to gene 1296, even though with a much lower rank in terms of the marginal effect on the response. It reiterates our standpoint that our proposed methods can jointly assess the significance of relevant features.


\begin{figure}[H]
  \centering
  \includegraphics[width=1\textwidth,height=0.45858\textwidth]{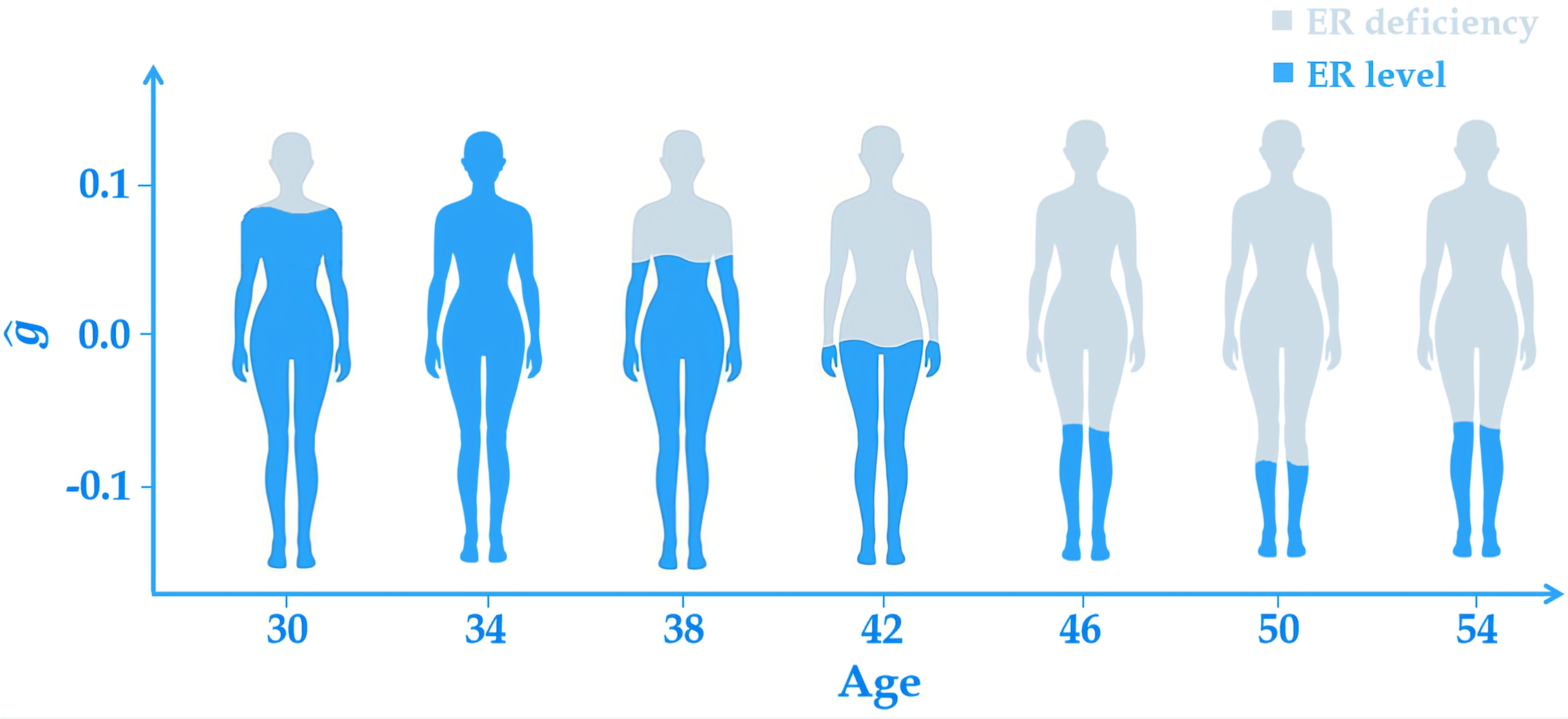}
  \vspace{-10mm} \caption{The cartoon illustration of {$\wh{g}$} indicating ER levels changes with age, {which is obtained by (\ref{2.6}) using $k=23$ candidate genes screened by SPLS.}}
  \label{fitted nonlinear function}
  
\end{figure}

Lastly, we present the estimated nonparametric function curve in Figure \ref{fitted nonlinear function}, which displays the trend of ER levels changing with patients' age in a cartoon manner. {Specifically, along with the seletion step mentioned above, we use the same $k=23$ candidate genes screened by SPLS to estimate $g$ by (\ref{2.6}).} It shows that the value of effect first increase during one's youth (from age 30 to 34), subsequently decrease in middle age (from age 34 to 50), and then slightly bounce back when growing old (from age 50 to 54). This result is consistent with many authoritative studies in medicine, such as \cite{CG2004}, 
further underscoring the validity of the proposed method.

Hence, we show that our proposed methods have significance performances in controlling FDR for high-dimensional partially linear models from a practical point of view.

\section{Conclusion}
\label{Conclusion}

This paper considers the problem of variable selection for the partially linear model with FDR control by using generalized knockoff features.
Incorporating selection probability as feature importance scores, we develop a Stab-GKnock procedure.
The finite-sample FDR control and asymptotically power results are established for the proposal under some regularity conditions. 
A two-stage procedure based on joint screening is also developed under high dimensionality.
There are some future directions. 
An interesting direction is to investigate the applicability of Stab-GKnock and Sparse-PLS to other semiparametric models, such as varying-coefficient models and generalized semiparametric models. 
Moreover, the study of power analysis based on SPD statistics can be extended to other aspects in view of more complex data, such as Gaussian graphic model \citep{LM2021}, 
nonparametric adaptive model \citep{DLL2022}, 
and structure change detection \citep{LSK2022},
noting that they are not trivial and call for further theoretical results. 
In addition, another interesting issue is to extend our Stab-GKnock under model-X knockoff framework to study the robust knockoff-based method for high-dimensional semiparametric models with heavy-tailed error distribution or misspecified feature distribution, which may combine with the idea in \cite{FGL2023}. 


\section*{Acknowledgements}
     This research was supported by the National Natural Science Foundation of China 
     (12271046, 12101119, 11971001 and 12131006) and the Fundamental Research Funds for the Central Universities (310422113).

\section*{Declarations}
\textbf{Conflict of interest} 
The authors declare that they have no conflict of interest.

\bibliographystyle{apalike_revised}

\bibliography{Stab-GKnock_manuscript}

\end{document}